\documentclass[12pt]{iopart}
\pdfoutput=1
\usepackage{amsmath}
\usepackage{amssymb}
\usepackage{graphicx}
\usepackage{hyperref}
\usepackage{xcolor}
\usepackage{xspace,slashed}
\usepackage{qcircuit}
\usepackage[utf8]{inputenc}
\usepackage{subfigure}
\usepackage[numbers,sort&compress]{natbib}
\usepackage{chngcntr}
\usepackage{orcidlink}

\usepackage{array}

\newcommand{\ionq}{{\tt aria-1}}
\newcommand{\ibm}{{\tt ibm{\_}torino}}

\renewcommand{\vec}[1]{\boldsymbol{#1}}

\newcommand{\QBaff}{QuantumBasel, Schorenweg 44B, CH-4144
   Arlesheim, Switzerland.}
\newcommand{\UniBas}{Center for Quantum Computing and Quantum Coherence (QC2), Department of
Physics, University of Basel, Klingelbergstrasse 82, CH-4056 Basel, Switzerland}

\begin{document}

\title[\footnotesize Cross-platform benchmark of style-qGAN for data augmentation]{Cross-platform hardware benchmark of style-based quantum GANs for data augmentation on superconducting and trapped-ion processors}

\author[Julien Baglio]{Julien Baglio$^{1,2}$\,\orcidlink{0000-0001-6691-6108}\newline}
\address{$^1$\QBaff\newline}
\address{$^2$\UniBas}
\ead{julien.baglio@quantumbasel.com}

\maketitle

\begin{abstract}
  In the noisy intermediate-scale quantum era, controlled benchmarks of quantum machine-learning workloads across hardware modalities are needed to quantify how given algorithms behave under native provider execution stacks. This work presents such a benchmark for the style-based quantum generative adversarial network (qGAN) on a high-energy physics data-augmentation task. We compare two commercially available gate-model quantum computers: the IBM \ibm\ hardware, based on superconducting transmon qubits from the Heron chip and the IonQ \ionq\ hardware, based on trapped-ion qubits. The generator architecture and trained parameters are kept fixed, and built-in mitigation is disabled when possible. We report quality and runtime metrics under each provider's native stack. The workflow uses circuit replication across available qubits, up to 48 on IBM and 24 on IonQ, to reduce the number of submitted jobs required for the target sample set. To our knowledge, this is one of the first controlled style-based qGAN hardware-to-hardware comparisons for this data-augmentation task. We observe that both platforms complete the task successfully, with marginal Kullback-Leibler divergences somewhat lower on \ionq, while end-to-end runtime is significantly shorter on \ibm. These results are an application-specific tradeoff benchmark, not a claim of algorithmic novelty.
\end{abstract}


\section{Introduction}

While quantum computing has seen great progress happening in the past few years, and is now focusing more on use-case studies and applications, the commercially available quantum computers accessible via the cloud are still noisy intermediate-scale quantum (NISQ) computers. They are heterogeneous in connectivity, gate speeds, gate and readout fidelity, and software stacks. Complementing algorithm advancements, controlled cross-platform benchmarks of fixed workloads isolate platform-dependent quality and throughput effects that matter for near-term deployment~\cite{cerezo2021variational,bharti2021noisy,dalzell2023quantum}.

Amongst quantum algorithms, quantum machine learning (QML)~\cite{biamonte2017quantum,schuld2018supervised} is of high interest given the range of real-world applications~\cite{gujju2023quantum}. We focus on quantum generative models~\cite{benedetti2019generative,hamilton2019generative,coyle2020born}, in particular quantum generative adversarial networks (qGANs)~\cite{dallaire2018quantum,lloyd2018quantum,zoufal2019quantum,hu2019quantum,situ2020quantum,romero2021variational,BravoPrieto2022,Chaudhary2023}, for a survey see Ref.~\cite{electronics12040856}. These models are instances of parameterized quantum circuits~\cite{benedetti2019parameterized,Larocca2023,barthe2024expressivity}. Trainability of qGANs has been studied~\cite{rudolph2023trainability,letcher2024tight}, in particular Ref.~\cite{letcher2024tight} indicates that qGANs with shallow quantum generators suffer less from barren plateaus, making them promising for generative applications. We are in particular interested in data augmentation: A model is trained on a small set of input samples and learns to sample the underlying distribution. This arises across domains, e.g.\ finance~\cite{Zhu2022}, healthcare~\cite{Sandfort2019}, and drug discovery~\cite{li2021quantum}.

The quantum algorithm we have selected for data augmentation is the style-based qGAN which we have proposed with other colleagues in Ref.~\cite{BravoPrieto2022}. In the style-based qGAN, the latent variables of the quantum generator are repeatedly encoded over the entire quantum network and not only in the first quantum gates. This approach has been shown to generalize the standard qGANs~\cite{zoufal2019quantum,situ2020quantum,romero2021variational}. The data augmentation was demonstrated on real-world data provided by Monte Carlo event distributions typically encountered in particle physics at hadron colliders and which display highly non-Gaussian profiles~\cite{BravoPrieto2022}. Not only is this type of distributions a difficult playground for classical GANs in the context of data augmentation, but it has also been shown that significantly smaller Kullback-Leibler (KL) divergences are achieved by the style-based qGAN compared to the standard qGAN.

The data augmentation task in Ref.~\cite{BravoPrieto2022} was demonstrated on IBM superconducting-qubit quantum hardware~\cite{Bravyi2022} and hinted at on an IonQ trapped-ion hardware~\cite{Bernardini2024,2019Wright} but not thoroughly benchmarked across modalities. These two hardware families are widely used for QML tasks, and we therefore select IBM and IonQ cloud systems for a controlled cross-platform benchmark of style-based qGAN data augmentation. Other vendors also offer superconducting (for example Rigetti Ankaa class processors~\cite{Rigetti2021}) or trapped-ion machines (for example the Helios Quantinuum~\cite{2021Pino} system). IBM's Heron-class processors, with up to 156 qubits, and IonQ's more recent generations~\cite{chen2023benchmarking}, with 36 qubits, differ in typical connectivity, gate timing, and gate and readout error budgets. A controlled hardware benchmark can indicate which modality successfully runs this workflow and which specifications most limit quality or throughput. For shallow variational circuits such as the style-based qGAN generator, gate and readout errors are often more salient than raw coherence times. Provider-native error mitigation and suppression tools exist but are deliberately excluded from our main comparison to keep the benchmark closer to ``bare'' execution, see below.

This work has two goals:
\begin{enumerate}
\item Assess quantitatively whether a fixed style-based qGAN generator can perform good data augmentation on two different hardware architectures, superconducting and trapped-ion systems. This was hinted in Ref.~\cite{BravoPrieto2022} but a quantitative study is still lacking.
\item Compare quality/runtime tradeoffs for this specific workload. The contribution is a controlled experimental benchmark, in the same spirit as cross-hardware benchmark studies such as Ref.~\cite{LOTSTEDT2024140975}. It is not a claim of a new qGAN theory or a new training algorithm. Because we want to compare native provider stacks as directly as possible, we do not apply additional user-level error suppression or error mitigation in the main hardware results. Such a study is important but goes beyond the scope of this work. We also revisit the practical implementation of the style-based qGAN in particular to improve sample throughput. Quantum multi-programming and qubit mapping for concurrent jobs on a single device have been studied, for example in QuCloud~\cite{Liu2021QuCloud} and in the {QuMC} framework~\cite{Niu2023QuMC}, however this work does not address that problem class; it simply uses circuit replication.
\end{enumerate}

Concerning classical baselines, we emphasize that this work addresses a different question than algorithmic quantum-versus-classical benchmarking. The style-based qGAN algorithm was introduced previously in Ref.~\cite{BravoPrieto2022}, and comparisons against classical pipelines have already been discussed in the literature for other use cases based on this quantum architecture~\cite{chang2024}. Here we intentionally keep the focus on a hardware-level benchmark, where the same style-based qGAN workflow is executed on two quantum platforms with a consistent protocol, to isolate platform-dependent quality and runtime tradeoffs for this application. Compared to Ref.~\cite{BravoPrieto2022}, we also update the implementation stack and use replication of the base ansatz across available qubits. This allows us to perform runs using up to 24 qubits on the IonQ \ionq\ system and up to 48 qubits on the IBM \ibm\ system. This leads to a substantial reduction in the number of submitted jobs needed for a fixed number of generated samples, which in turn reduces the end-to-end execution time. We present this replication as an execution strategy used in this benchmark study.

To make the interpretation of this benchmark explicit, we state here the main choices for our protocol, indicating the benchmark scope, the fairness criteria, and the limitations of our study.
\begin{itemize}
\item The quantum generator architecture and trained parameters are fixed across hardware platforms. We do not retrain separately on each backend and we use the checkpoints from Ref.~\cite{BravoPrieto2022} to also allow for a direct comparison with the results from that work.
\item We compare native hardware execution and provider software paths. This captures user-relevant practical performance but does not isolate all low-level hardware factors.
\item The shot counts differ across platforms. On IBM, we use the nominal shot budget of the system; on IonQ, we choose a smaller shot count guided by the noise-model study in \ref{sec:appendixnoise} to limit wall-clock time. This is a pragmatic workflow choice and does not by itself establish a fully apples-to-apples statistical budget across platforms. Runtime figures are therefore reported both as end-to-end totals and as normalized per-shot/per-circuit quantities.
\item Main hardware results correspond to one run per platform due to limited quantum resources. The uncertainty bands on KL metrics are estimated from sample-variance propagation.
\item No additional user-level error mitigation is applied in the main comparison, as explained above.
\end{itemize}
These choices imply that our conclusions are application-specific and workload-specific, and should not be interpreted as universal rankings of superconducting versus trapped-ion platforms.

This paper is organized as follows. In Section~\ref{sec:technology} we present the two quantum computers on which we implement the style-based qGAN algorithm: the \ibm\ device based on the IBM Heron superconducting transmon qubits and the \ionq\ device based on the IonQ trapped-ion qubits. In Section~\ref{sec:algorithm}, we summarize briefly the design and architecture of the style-based qGAN. We also introduce circuit replication, which reduces the number of effective runs required to obtain the total number of samples from the quantum generator, and we contrast it with general-purpose multi-programming as discussed above. In Section~\ref{sec:results} we present the experimental results of the hardware implementation in the context of data augmentation using Monte Carlo distributions for high-energy physics as real-world data, comparing the performance of \ionq\ and of \ibm. In Section~\ref{sec:conclusion} we present our conclusion and outlook for further exploration.
In \ref{sec:appendixnoise} we include a noise simulation for the IonQ \ionq\ device comparing runs using 512 shots and 1024 shots, while \ref{sec:appendixibm} presents additional results on the IBM {\tt ibm{\_}cusco} device based on the Eagle chip.

\section{Superconducting transmon and trapped-ion quantum computers}
\label{sec:technology}

We have implemented the style-based qGAN algorithm on two different hardware technologies, namely superconducting transmon qubits as manufactured by IBM and trapped-ion qubits as manufactured by IonQ.

\begin{figure}[t!]
  \centering
  \includegraphics[width=0.48\textwidth]{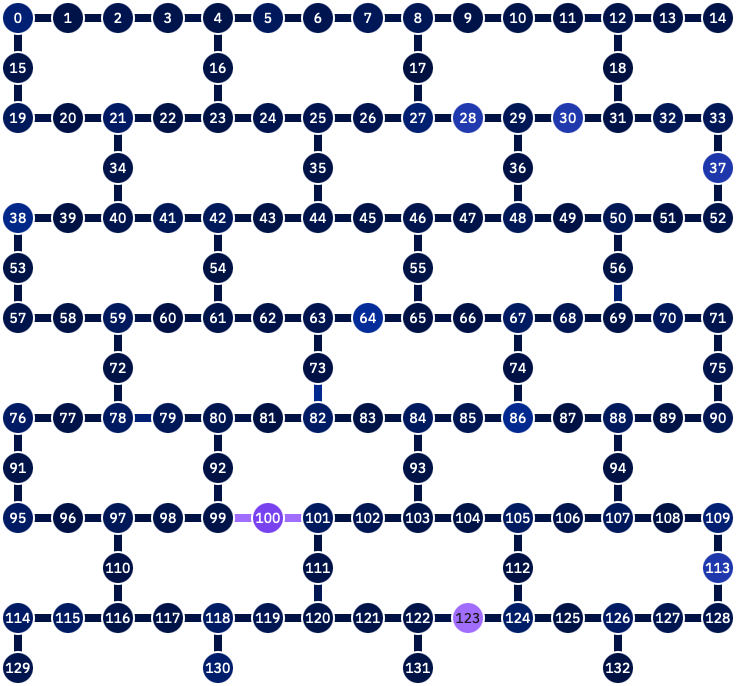}
  \caption{Qubit coupling map of the 133-qubit \ibm\ system based on the IBM Heron chip.\label{fig:ibmtorino}}
\end{figure}

IBM uses superconducting transmon qubits~\cite{Koch2007,Krantz2019} for its quantum chip. The qubit is made out of a Josephson junction of the size of about 100 nm~\cite{Fritz2019}, acting as a nonlinear inductor, coupled with a shunting capacitor. This circuit creates an anharmonic oscillator. The qubits are cooled down to mK temperatures to enter the superconducting regime and exhibit quantized energy levels. The typical resonance frequency is around 5 GHz. Controlling of the qubits, including readout as well as gate operations, is performed thanks to microwave resonators coupled to the chip~\cite{Chow2011,Jeffrey2014}. We have used the \ibm\ quantum system which is the latest IBM prototype using the Heron architecture with 133 qubits. The qubit connectivity is defined by a heavy-hex lattice\footnote{See this \href{https://www.ibm.com/quantum/blog/heavy-hex-lattice}{IBM blog post} on heavy-hex lattice architecture.} and the Heron chip, which is the baseline for the new IBM System 2 system, is a substantial improvement over the previous Eagle chip with ten times better two-qubit gate fidelity\footnote{See the different \href{https://docs.quantum.ibm.com/run/processor-types}{IBM quantum processor types} in the IBM documentation.}. The qubit coupling map for the \ibm\ quantum system using the Heron chip is displayed in Figure~\ref{fig:ibmtorino} and the most important parameters from the technical specifications of the device are listed in Table~\ref{table:specifications}.

Trapped-ion quantum computers use ionized atoms trapped with a laser~\cite{Bernardini2024}. IonQ quantum systems are currently based on ytterbium ions~\cite{2019Wright}, $^{171}\text{Yb}^+$, confined on a chip called a linear ion trap, featuring around 100 electrodes producing the rapidly oscillating electromagnetic fields needed for the trap. The hyperfine transitions of the $^2S_{1/2}$ ground state are used as $|0\rangle$ and $|1\rangle$ qubit states. The ions are cooled by the lasers using a combination of Doppler and resolved sideband cooling. The initial state is prepared in the ground state $|0\rangle$ via optical pumping. One- and two-qubit gate operations are processed sequentially, using 355nm-pulsed Raman beams, SK1 pulses for one-qubit operations~\cite{Merrill2012} and M{\o}lmer-S{\o}rensen interactions for two-qubit operations~\cite{Molmer1999,Sorensen2000}. The readout is performed on all ions at once, using a 369-nm laser resonant with the $^2 S_{1/2}\to  {^2} P_{1/2}$ transition. The laser operations allow for an all-to-all connectivity of the qubits. We have used the IonQ \ionq\ quantum system, with 25 qubits and the technical specifications as listed in Table~\ref{table:specifications} for the most important parameters.

\begin{table}[t!]%
  \centering
  \footnotesize
  \begin{tabular}{l|cc}
    & \ibm
    & \ionq \tabularnewline
      \hline
      \# of qubits
    & $133$
    & $25$ \tabularnewline
      Coherence time $T_1$ ($\mu$s)
    & $157$
    & $10^8$ \tabularnewline
      Coherence time $T_2$ ($\mu$s)
    & $140$
    & $10^6$ \tabularnewline
      One-qubit gate time ($\mu$s)
    & $0.032$
    & $135$ \tabularnewline
      Two-qubit gate time ($\mu$s)
    & $0.101^a$
    & $600^b$ \tabularnewline
      Readout time ($\mu$s)
    & $1.56$
    & $300^c$ \tabularnewline
      One-qubit gate error rate
    & $5.8\times 10^{-4}$
    & $3\times 10^{-4}$\tabularnewline
      Two-qubit gate error rate
    & $5.3\times 10^{-3}$
    & $6\times 10^{-3}$\tabularnewline
      Readout error rate
    & $2.7\times 10^{-2}$
    & $5.1\times 10^{-3}$\tabularnewline
    \hline
  \end{tabular}\\%
  $^a$CZ gate; $^b$M{\o}lmer-S{\o}rensen gate; $^c$On all qubits at once.%
  \normalsize
  \caption{\label{table:specifications} List of the most important parameters from the technical specifications of the IBM \ibm\ (left) and IonQ \ionq\ (right) devices (average values over all the qubits). The actual values change over time after each calibration of the system and reflect the specifications at the time of our experiments (for the IBM system: December 13th, 2023: for the IonQ system: spanned mid-February to mid-March 2024).}
\end{table}

It is quite instructive to compare the parameters in Table~\ref{table:specifications} for the two quantum computers \ionq\ and \ibm. Besides the very important difference in terms of connectivity, with all-to-all for the trapped-ion architecture vs the heavy-hex lattice for the superconducting transmon qubit architecture, the other salient differences lie in the coherence time, the one-qubit gate error and the readout error.

Trapped-ion qubits have a very long coherence time, $T_1=100$~s, compared to the typical coherence time of order 200~$\mu$s for superconducting transmon qubits. This has an impact on the number of gate operations that can be performed before measuring just noise: IonQ systems can run in principle deeper circuits, as evidenced by the T2-over-two-qubit-gate-time ratio which is of order 1400 on \ibm\ and of order 1700 on \ionq. The higher this ratio is, the more operations you can perform in a given quantum circuit before losing the whole quantum coherence. On the other hand, operations on \ionq\ take a much longer time compared to operations on \ibm, as there is a three-order- of-magnitude difference in the two-qubit gate operations time.

The two-qubit gate error rates are quite comparable, while the IonQ \ionq\ device has a significantly better readout fidelity than the IBM \ibm\ device, by one order of magnitude. We expect this readout fidelity to have more impact on the accuracy of our data augmentation experiment than the difference in coherent time between the two quantum devices, as our quantum circuits are quite shallow.

\section{Style-based quantum GAN and circuit replication}
\label{sec:algorithm}

\subsection{Workflow of a (quantum) GAN}

\begin{figure}[t!]
  \centering
  \includegraphics[width=0.45\textwidth]{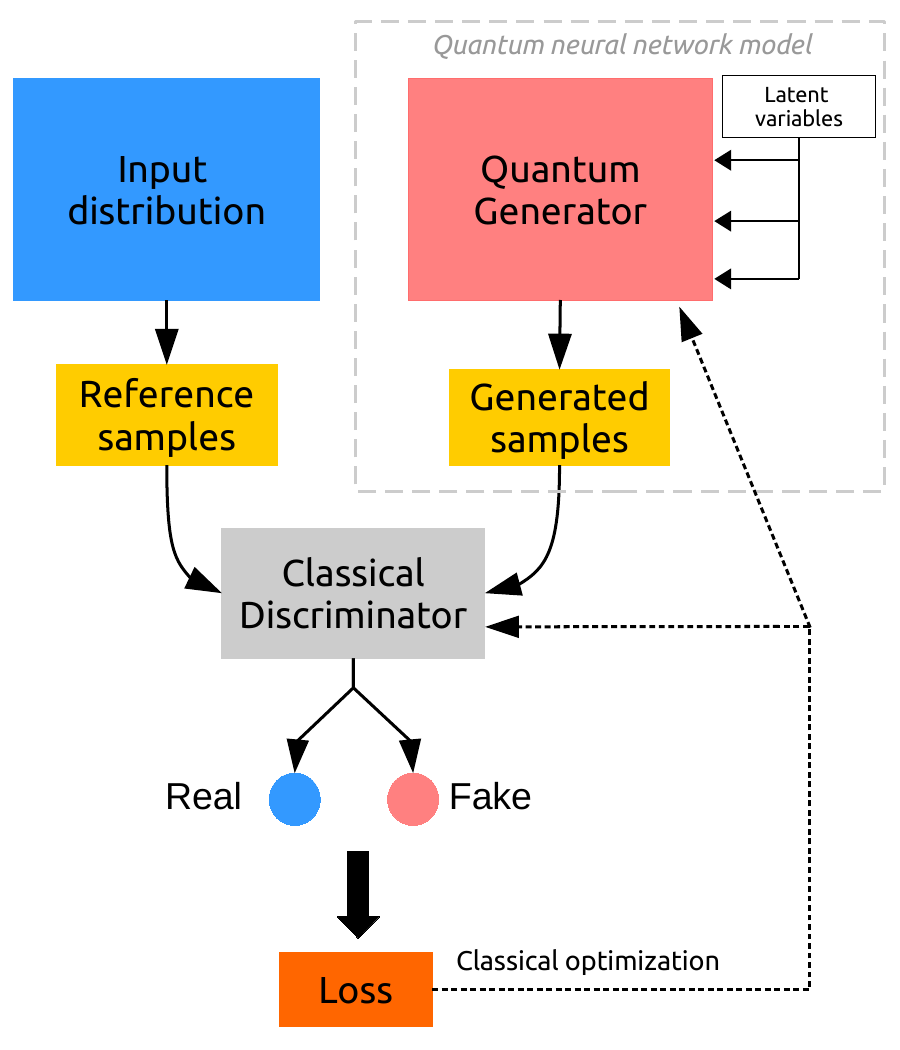}
  \caption{\label{fig:scheme} Workflow of a qGAN training where the discriminator uses a classical neural network while the generator is a quantum network. This is the setup of the style-based qGAN algorithm we use in this work. The generated samples out of the quantum generator are compared against the reference samples, the latter being taken from the input training data. Both generated and reference samples are used to train the discriminator, using an appropriate loss function for both the discriminator and the generator. The parameters of both networks are then updated and the procedure is repeated in an adversarial approach until the desired precision is reached. The figure is taken from Ref.~\cite{BravoPrieto2022}.}
\end{figure}

We briefly present in this section how a GAN, be it classical or quantum, is implemented. A GAN contains two competing networks, the generator and the discriminator, which are trained alternately following an adversarial procedure~\cite{goodfellow2014generative}. The goal of the generator is to produce candidate data (or ``fake data'') out of random noise input, while the goal of the discriminator is to distinguish the candidate data produced by the generator from the training data it is fed with (the ``real data'').

The training procedure follows a zero-sum two-player game until (ideally) a Nash equilibrium is reached: the discriminator cannot discriminate  between fake and real data better than randomly selecting true or false, and the parameters of the generators have now reached their final values so that the random input distribution of the generator is converted into realistic data by the network. The step from a classical GAN to a quantum GAN (qGAN) can be realized either by using a quantum architecture for the generator, or the discriminator, or both~\cite{dallaire2018quantum,electronics12040856}. Note that game-theory arguments indicate that a Nash equilibrium may not exist in all cases for GANs~\cite{farnia2020gans}, related to the fact that the training of classical GANs may be challenging with, amongst other issues, vanishing gradients~\cite{moghadam2022game}. Interestingly this is another argument in favor of qGANs which, for example, are more likely to avoid the problem of vanishing gradients as indicated in Ref.~\cite{letcher2024tight} when the generator of the GAN is a quantum network.

The style-based qGAN algorithm uses the hybrid approach with a classical discriminator network and a quantum architecture for the generator, as illustrated in Figure~\ref{fig:scheme}. As explained in Ref.~\cite{BravoPrieto2022}, using a classical discriminator leads to a faster convergence of the loss function in our case. We also stress that the quantum generator is what we eventually want to run extensively after the training: The discriminator network is not useful in the deployment as the target is the generation of synthetic data thanks to the generator, for which we are interested in the possibility of performance improvements using a quantum device.

The classical discriminator is composed of a deep convolutional neural network with four convolution layers. The exact details of its implementation can be found in the code~\cite{stavros_efthymiou_2024_10932110} which is based on the open-source Python quantum software development kit  {\tt qibo}~\cite{efthymiou2020qibo}.

\subsection{Quantum generator of the style-based qGAN and training procedure}
\label{sec:sec:training}

\begin{figure*}[t!]
  \centering
  \includegraphics[width=0.65\columnwidth]{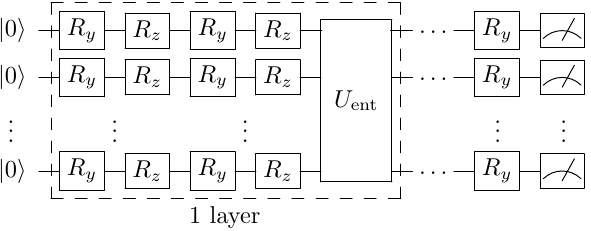}
  \caption{\label{fig:ansatz} Ansatz for the quantum generator of the style-based qGAN. $U_{\rm ent}$ in our case stands for controlled $R_z$ rotations.The figure is taken from ~\cite{BravoPrieto2022}.
  }
\end{figure*}

The quantum generator of the style-based qGAN is depicted in Figure~\ref{fig:ansatz}. It is essentially a quantum neural network for a quantum feature map encoding the latent vector (the random noise with a given latent dimension, generated from a standard Gaussian noise distribution) into a quantum state $|\Psi\rangle$. We then perform on this quantum state the measurement of an expectation value to obtain the fake samples. Each layer of this quantum feature map is comprised of a set of alternating $R_y$ and $R_z$ one-qubit gates followed by a set of entangling gates, that we chose as controlled rotations c-$R_y$ as explained in Ref.~\cite{BravoPrieto2022}. For our data augmentation experiment we use the same exact hyperparameters as in our previous work: one layer and a latent dimension of five for the latent vector $\vec{r}$. The gate angles follow a simple affine function of both the latent vector and the trainable parameters $\vec{p}$, such that we have
\begin{eqnarray}
  R^{l,m}_{y/z}(\vec{p}, \vec{r})  = R_{y/z}(p^{(l)} r^{(m)} + p^{(l-1)}),
\end{eqnarray}
with $m$ running from 1 to $D_{lat}=5$, the latent dimension, and $l$ running from 1 to the total number of trainable parameters, depending on the number of layers and on the number of qubits. This encoding is the quantum equivalent of classical machine learning encoding with weights and biases. The salient feature of the styled-based approach is the encoding of the latent variables all over the network, in a data re-uploading approach.

The last layer of the quantum circuit contains the measurement operator. The final sample $x\in \mathbf{R}^N$, where $N$ is the number of distributions (equivalent to the number of qubits in our data augmentation experiment for the base circuit), is generated as a vector containing the expectation values of individual Pauli $Z$ operators for each qubits over the final state $|\Psi(\vec{r})\rangle$ obtained with the circuit of Figure~\ref{fig:ansatz},
\begin{eqnarray}
  \vec{x} = \left(-\left\langle\sigma_z^1\right\rangle,-\left\langle\sigma_z^2\right\rangle,\hdots,-\left\langle\sigma_z^N\right\rangle\right)\,.
  \label{eq:samples}
\end{eqnarray}
We use a minmax pre-processing on the input data, so that the data for each distribution is rescaled within the range $[-1;1]$, using the power transform from the Python package {\tt scikit learn}~\cite{scikit-learn}. The data points generated by Equation~\ref{eq:samples} are post-processed through the reverse power transform to obtain the actual generated distributions.

The training procedure for the style-based qGAN is described at length in Ref.~\cite{BravoPrieto2022}. We briefly sketch it in order to introduce the functions we have used for the loss function and for the measure of performance of the quantum generator. We alternately train the discriminator network and the quantum generator network in an adversarial game: The discriminator is improved to distinguish the input (real) training data from the fake data produced by the generator, and then the generator is improved to produce better fake data to trick the discriminator. Both neural networks are trained with binary cross-entropy loss functions. The training is achieved when the Nash equilibrium of the two loss functions is reached:
\begin{eqnarray}
  \underset{\vec{p_g}}{\min}\,\,\mathcal{L}_G(\vec{p_g},\vec{p_d})  \,, \underset{\vec{p_d}}{\max}\,\,\mathcal{L}_D(\vec{p_g},\vec{p_d})  \,,
\end{eqnarray}
where $\mathcal{L}_G$ is the loss function of the generator, $\mathcal{L}_D$ is the loss function of the discriminator, $\vec{p_g}$ and $\vec{p}_d$ are the trainable parameters of the generator and the discriminator, respectively.

Our set of real data for the data augmentation experiment is based on high-energy physics, in particular Monte Carlo event generation for the Large Hadron Collider (LHC) at CERN. We use the same dataset as in Ref.~\cite{BravoPrieto2022}, in which the style-based qGAN was proposed, so that we can also compare our data augmentation experiment to the results we obtained before on a different IBM chip. The training data is composed of $10^4$ events for the production process $p p\to t\bar{t}$ at a 13 TeV center-of-mass energy, the production of top-antitop quarks at the LHC in the collision of two protons. We use the computer program MadGraph {\tt MG5\_aMC}~\cite{Alwall:2014hca,Frederix:2018nkq} to produce the training data as well as the $10^5$ reference samples to which we compare our generated $10^5$ samples. We get a set of three distributions corresponding to physical quantities describing the process: Mandelstam variables $s$ and $t$ (both in giga-electronvolt squared, or GeV$^2$) and the rapidity $y$. It is important to note that the $s$ and $t$ distributions, in particular, are highly non-Gaussian distributions. In order to quantify the quality of the generator, we use the Kullback-Leibler divergence (KL)~\cite{kullback1951information}. For two given distributions $P(x)$ and $Q(x)$ of discrete samples $x$, the KL divergence quantifies how similar these two distributions are,
\begin{eqnarray}
  D_{\text{KL}}(P||Q) = \sum_{x}P(x)\log\left(\frac{P(x)}{Q(x)}\right)\,,
\end{eqnarray}
which is essentially the difference between the entropy of the distribution $P(x)$ and the cross entropy of $P(x)$ with $Q(x)$. When the two distributions are identical, $D_{\text{KL}} = 0$, so that the smaller the KL divergence is, the more similar the two distributions are.

\subsection{Circuit replication for sample-throughput scaling}
\label{sec:algorithm:parallelization}

Compared to Ref.~\cite{BravoPrieto2022} we have modified the implementation of the quantum generator in several ways. The Python code is now based on qiskit primitive functions instead of {\tt backend.run} calls and accesses the IBM quantum device via qiskit runtime services. Specifically, we use the Sampler primitive to perform the measurement of the Pauli $Z$ matrix for each qubit, then marginalize properly on each qubit for the base circuit to obtain the corresponding quantum-measurement output for each physical distribution. This output is then classically post-processed in order to get the final generated physical distributions $s$, $t$, or $y$ and their corresponding two-dimensional correlations. We have used qiskit 0.43.1~\cite{matthew_treinish_2023_8003781} for generating our results, including also the qiskit-ionq provider\footnote{\url{https://github.com/qiskit-community/qiskit-ionq}} for running on IonQ devices.

In order to exploit the larger qubit counts now available compared to the IBM device used in Ref.~\cite{BravoPrieto2022}, we replicate the $N$-qubit base circuit $m$ times within one full circuit submitted to the quantum computer. This replication reduces the number of submitted runs needed to produce a target sample count $k$ from $k$ to $k/m$, at the cost of using $m\times N$ qubits in one execution on the hardware.

\begin{figure*}[t!]
  \centering
  \includegraphics[width=\textwidth]{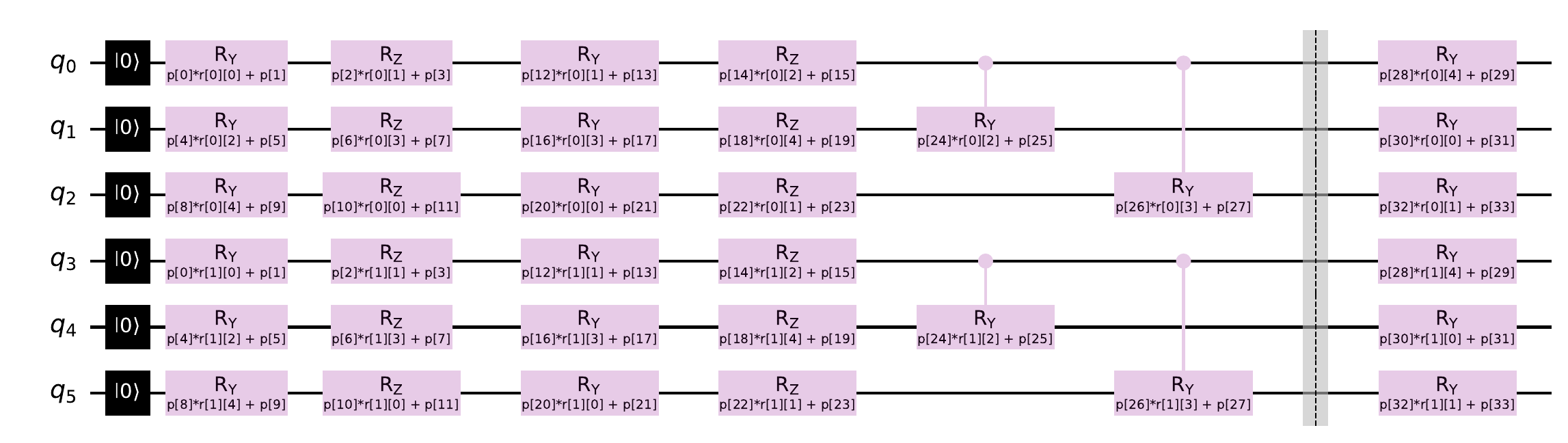}
  \caption{\label{fig:parallelization} Example of replication of a three-qubit base circuit of the style-based qGAN with two repetitions, using a total of six qubits. Only one layer is used with a (base) latent dimension of five. For simplicity the measurement operations have been omitted.}
\end{figure*}

An example is shown in Figure~\ref{fig:parallelization} for $m=2$ and $N=3$. For our default choice of $k=10^5$ target samples, two replicated blocks reduce the number of submitted runs to $k/m=50,000$. The number of trainable parameters remains that of the base three-qubit circuit, and the base latent dimension remains $D_{lat}=5$, while the total latent input size\footnote{The latent vector $\vec{r}$ has been promoted to a latent tensor $(r_{ij})$ for better readability, with $i$ being the index running over the number of replications and $j$ being the index over the base latent dimension $D_{lat}$.} becomes $m\times D_{lat} = 10$. In our experiments we have used 16 replications on \ibm, using 48 qubits, and eight replications on \ionq, using 24 qubits. For the same final sample count, $k=100,000$, it corresponds to 6,250 and 12,500 submitted runs, respectively. We have used the maximum number of qubits available on \ionq\ for our style-based qGAN experiment. It is in general expected that the higher the number of qubits used, the smaller the number of samples from the quantum circuit needed to generate the whole generated sample set, leading to a reduced time spent on the quantum device.

\subsection{Fairness implications of replication.}
Replication changes the wall-clock structure of the benchmark: larger $m$ reduces the number of circuit submissions but increases width and, on limited-connectivity devices, typically increases circuit depth after transpilation. Our IBM and IonQ runs use different replication factors ($m=16$ vs $m=8$) because of different available qubit counts, so the transpiled circuits are not identical across platforms. The trainable generator parameters are unchanged; only independent latent draws are supplied to each replicated block. Interpretation should therefore remain at the level of end-to-end workflow performance for this task, not as a claim that the underlying physical error channels are matched.

This replication is \emph{not} presented as a new scheduling or general-purpose multi-programming method. It is a deterministic tiling of the \emph{same} generator circuit to amortize job submission overhead and improve sample throughput for this benchmark. Cloud multi-programming for NISQ devices instead targets concurrent execution and mapping of \emph{distinct} jobs onto shared hardware, including qubit partitioning and crosstalk-aware allocation, as in QuCloud~\cite{Liu2021QuCloud} and the {QuMC} compiler framework~\cite{Niu2023QuMC}. Our benchmark does not implement or compare against such frameworks and co-scheduling unrelated circuits remains outside the scope of this work. It is, however, an interesting direction for future work.

\section{Data augmentation results and discussion}
\label{sec:results}

\subsection{Benchmark setup and cross-platform fairness}
We use the quantum generator of our style-based qGAN trained on $10^4$ Monte Carlo samples of the $p p\to t\bar{t}$ process at the LHC, as explained in Section~\ref{sec:sec:training}. In order to also compare to the previous implementation on the IBM {\tt ibmq{\_}santiago} device, which was using a five-qubit Falcon r4T chip, we have performed the data augmentation experiments on \ibm\ and on \ionq\ devices using the same set of trained parameters as in Ref.~\cite{BravoPrieto2022}. The quantum generator consists of one layer of the base three-qubit circuit displayed in Figure~\ref{fig:ansatz}, replicated over the whole circuit as explained in Section~\ref{sec:algorithm:parallelization} and presented in Figure~\ref{fig:parallelization}.

\textbf{What is aligned across platforms:} We use the same trained generator parameters, the same dataset and preprocessing, the same target sample count $k=10^5$, the same KL-based quality metrics, and no additional user-level mitigation in the main hardware comparison.

\textbf{What differs and is not fully matched:} (i)~Shot budgets: $n_{\rm shots}=4,000$ on IBM vs $512$ on IonQ. We explain below and in \ref{sec:appendixnoise} in detail the reasons behind this imbalance. (ii)~Replication factors and qubit footprint: $m=16$ on IBM vs $m=8$ on IonQ, hence different transpiled depths and readout structure. (iii)~Provider batching: IBM allows many circuits per job, while IonQ runs in our campaign do not. (iv)~Single hardware campaign per platform: because quantum resources are still scarce, we have only performed one data augmentation task per platform. As such, the uncertainty on the KL metrics is propagated from sample variance, not from independent hardware repetitions.

\textbf{How we report runtime:} We quote total wall-clock time to obtain the full sample set, plus per-circuit and per-shot times where available, so that throughput can be compared both at the workflow level and at the circuit level despite differing shot and batching choices.

\subsection{Quality of generated distributions}

Each run on the quantum hardware consists of a circuit execution to build the quantum state and then a number of measurements (shots) of the Sampler primitive to build the expectation values as displayed in Equation~\ref{eq:samples}. We use $n_{shots} = 4,000$ on IBM \ibm\ as this is the nominal number of shots on IBM systems. Note that this is a factor of four higher than the number of shots chosen on the IBM {\tt ibmq{\_}santiago} device in Ref.~\cite{BravoPrieto2022}. In principle, the higher $n_{shots}$ is, the less sensitive a quantum experiment is to the statistical error in building the expectation values out of the shots. As IBM systems allow for sending circuits in parallel, for a maximum of 300 circuits, we have performed 22 runs in total to obtain the full set of $10^5$ samples: two sessions, each consisting of ten runs with 300 circuits and one run with 125 circuits.

\begin{figure*}[t!]
  
  \hspace{1.2em}\subfigure[]{\includegraphics[width=0.30\textwidth]{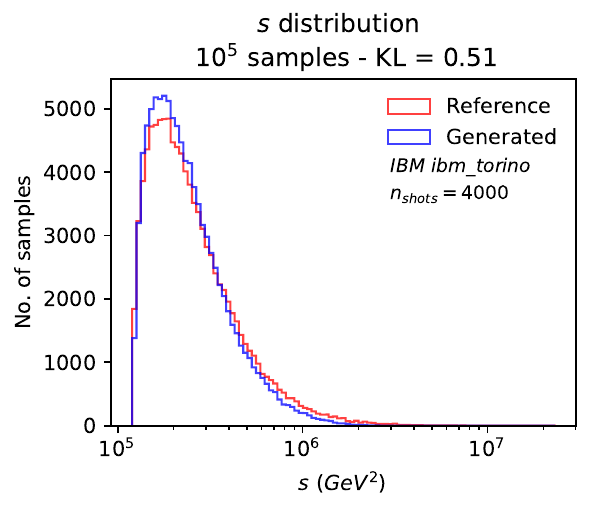}}%
  \subfigure[]{\includegraphics[width=0.30\textwidth]{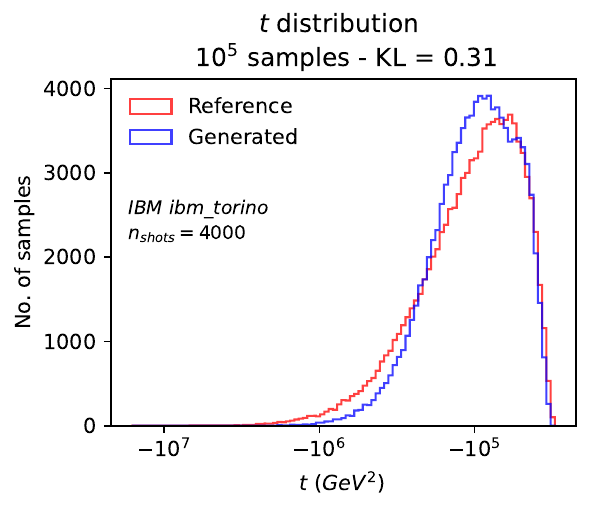}}%
  \subfigure[]{\includegraphics[width=0.30\textwidth]{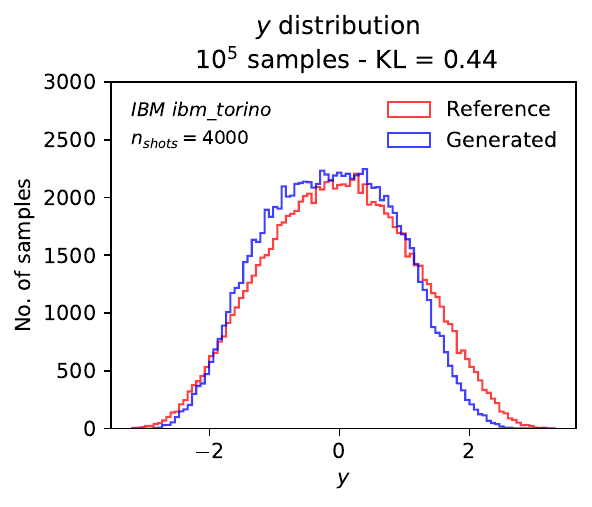}}

  \hspace{0.7em}\subfigure[]{\includegraphics[width=0.335\textwidth]{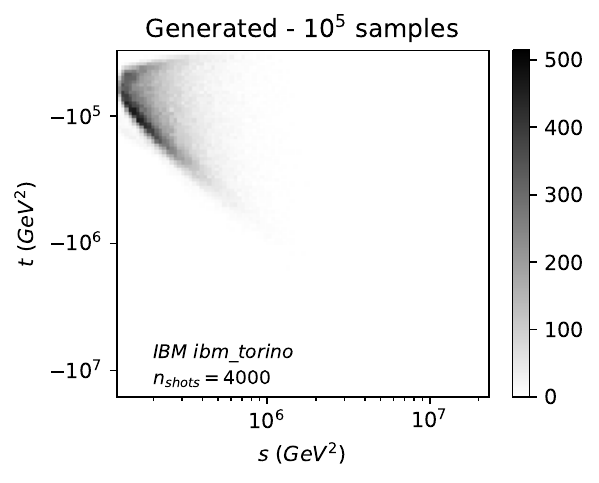}}%
  \subfigure[]{\includegraphics[width=0.310\textwidth]{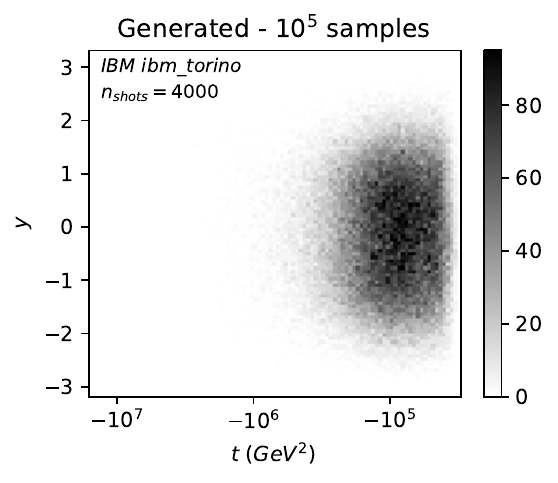}}%
  \subfigure[]{\includegraphics[width=0.325\textwidth]{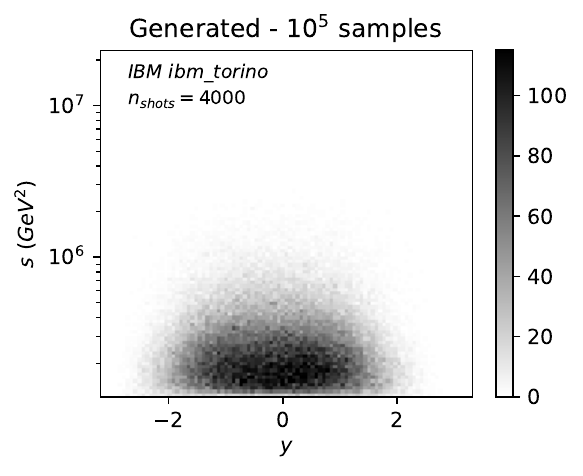}}

  \hspace{0.8em}\subfigure[]{\includegraphics[width=0.325\textwidth]{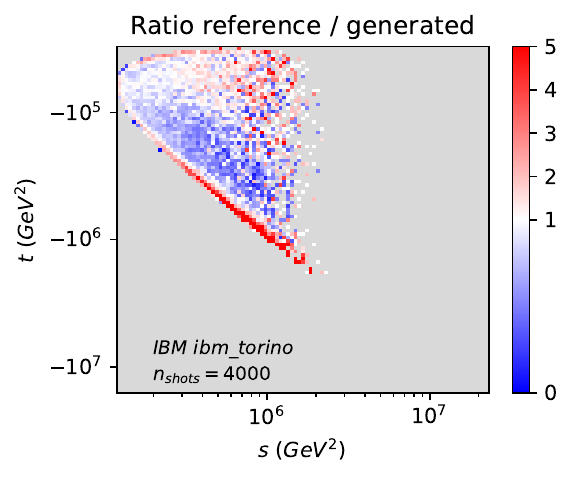}}%
  \subfigure[]{\includegraphics[width=0.305\textwidth]{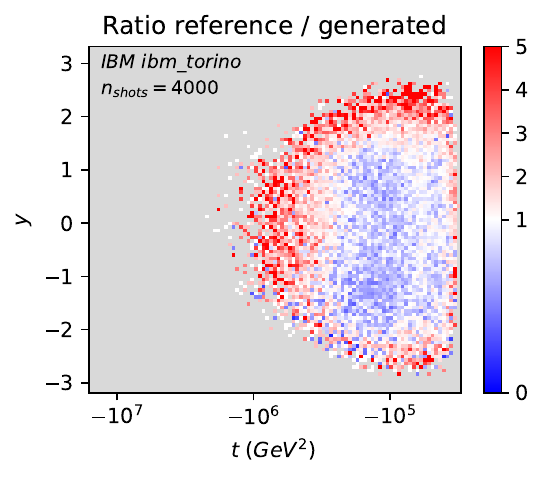}}%
  \subfigure[]{\includegraphics[width=0.32\textwidth]{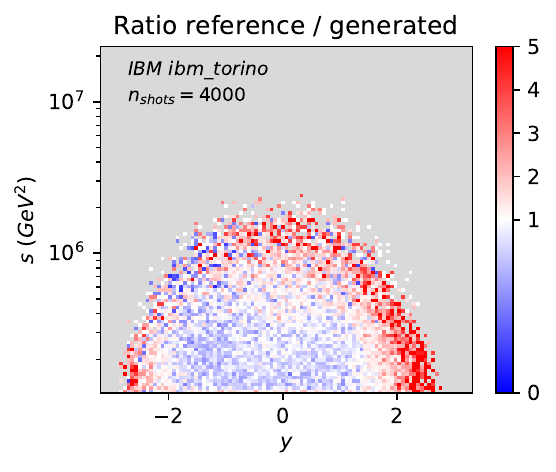}}

  \caption{\label{fig:ibmtorinoresults}
    Results for the data augmentation experiment on the IBM \ibm\ device using the Heron chip, using the style-based qGAN generator  trained with $10^4$ samples of Monte Carlo data for the physical observables $s,t,y$ in $pp\rightarrow t\bar{t}$ production at the LHC.
    Panels (a)--(c): marginal sample distributions for $s$, $t$, and $y$, respectively.
    Panels (d)--(f): corresponding two-dimensional sampling projections.
    Panels (g)--(i): ratio to the reference underlying prior Monte Carlo distribution. A gray background is used to highlight ratio-one regions, shown in white.}
\end{figure*}

The amount of time needed for any gate operation on trapped-ion quantum computers is larger than the corresponding timing on superconducting transmon qubits systems as exemplified in Table~\ref{table:specifications}. It is therefore required to chose $n_{shots}$ as small as possible on \ionq, without degrading the quality of the output, while keeping in mind that in general it is possible to chose a smaller number of shots than on superconducting transmon qubits systems because the quality of the qubits is on average higher on trapped-ion computers. We have performed a noise simulation study using \ionq\ noise model provided by IonQ with two different values for the number of shots: $n_{shots}=512$ and $n_{shots}=1024$. The results are presented in \ref{sec:appendixnoise} and they lead us to chose $n_{shots}=512$ for our data augmentation experiment on \ionq. We have performed a total of $12,500$ runs on the IonQ \ionq\ device as parallel circuit execution is not (yet) available on IonQ systems.

\begin{figure*}[t!]
  
  \hspace{1.2em}\subfigure[]{\includegraphics[width=0.30\textwidth]{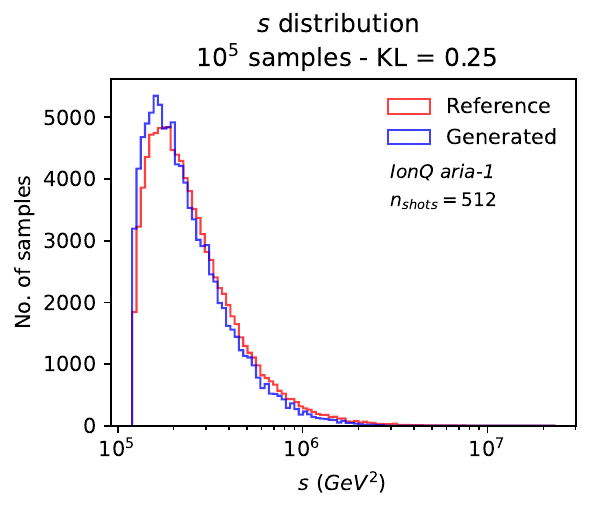}}%
  \subfigure[]{\includegraphics[width=0.30\textwidth]{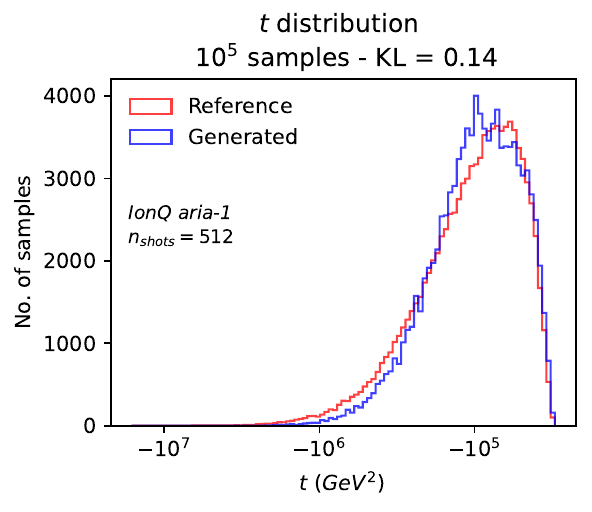}}%
  \subfigure[]{\includegraphics[width=0.30\textwidth]{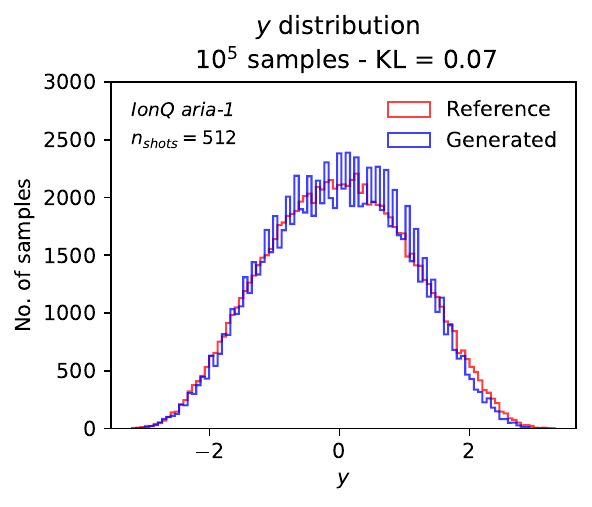}}

  \hspace{0.7em}\subfigure[]{\includegraphics[width=0.335\textwidth]{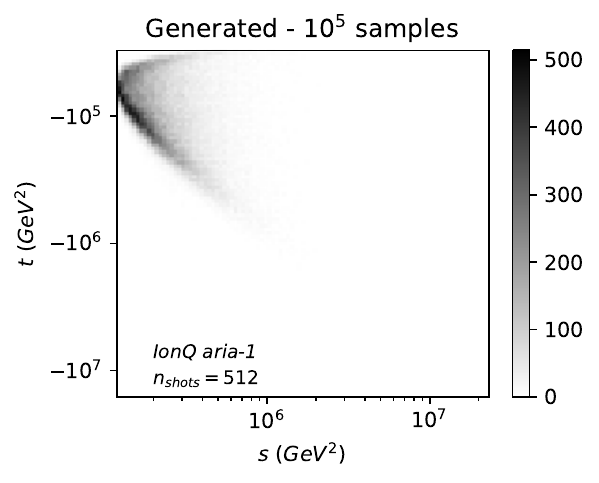}}%
  \subfigure[]{\includegraphics[width=0.310\textwidth]{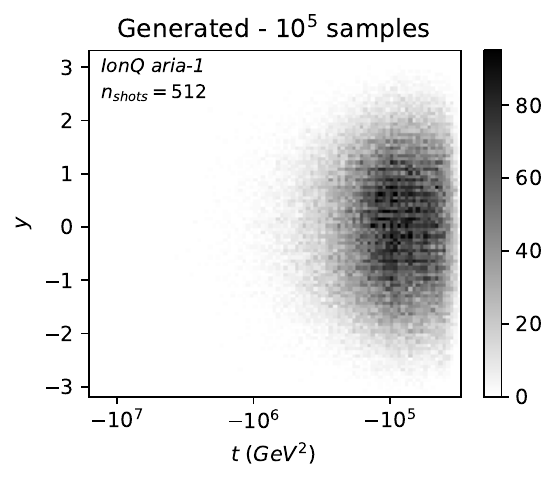}}%
  \subfigure[]{\includegraphics[width=0.325\textwidth]{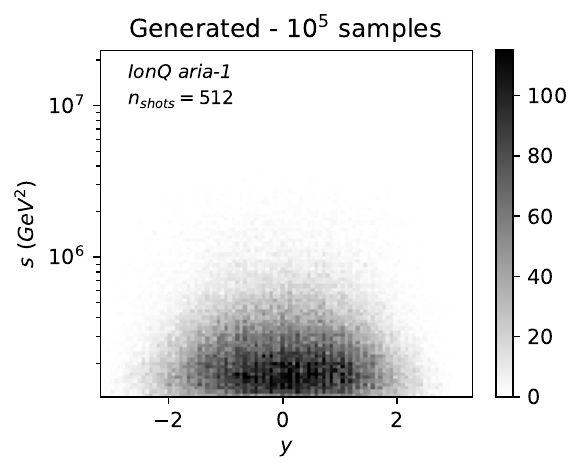}}

  \hspace{0.8em}\subfigure[]{\includegraphics[width=0.325\textwidth]{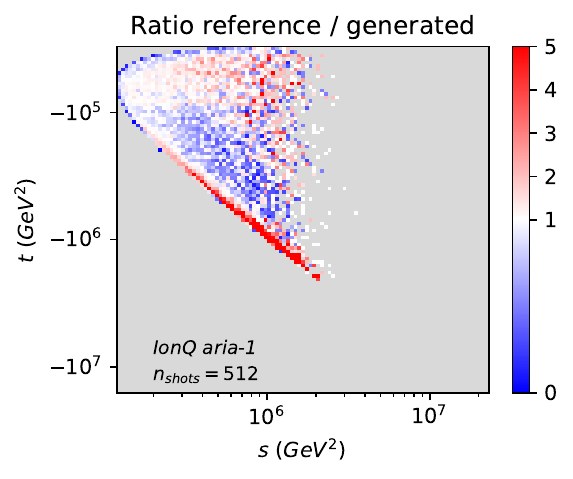}}%
  \subfigure[]{\includegraphics[width=0.305\textwidth]{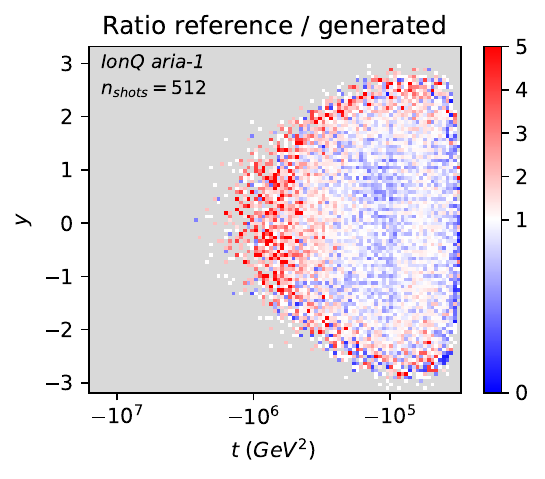}}%
  \subfigure[]{\includegraphics[width=0.32\textwidth]{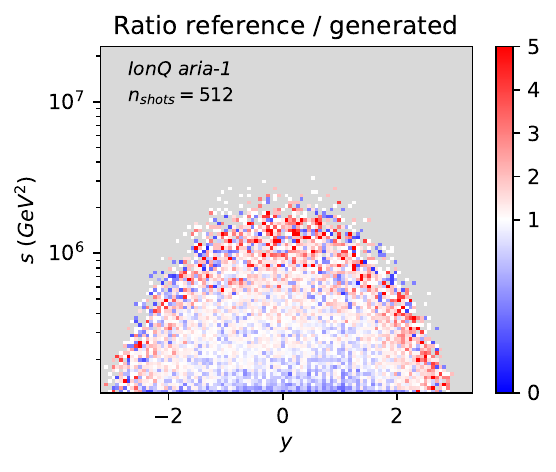}}

  \caption{\label{fig:ionqaria1results}
    Same panel organization as in Figure~\ref{fig:ibmtorinoresults}, but for the experiment on the IonQ \ionq\ device.
    Panels (a)--(c): marginal sample distributions for $s$, $t$, and $y$, respectively.
    Panels (d)--(f): corresponding two-dimensional sampling projections.
    Panels (g)--(i): ratio to the reference underlying prior Monte Carlo distribution.}
\end{figure*}

We display in Figure~\ref{fig:ibmtorinoresults} the results of the data augmentation on the IBM \ibm\ device, with a grid of 100 linearly spaced bins for $y$ and 100 log-spaced bins for $s$ and $t$. We will use this binning in all of our results. The replicated circuit built out of the base circuit of Figure~\ref{fig:ansatz} is transpiled to the \ibm\ device, which is a step that adds more gates in the circuits as: 1) there is no all-to-all connectivity on the Heron chip, resulting in the insertion of swap gates to connect some of the qubits; 2) the circuit has to be adapted to the set of native gates of the quantum device. In the top row, we compare the one-dimensional projections of samples generated by the quantum generator of the style-based qGAN with the reference input distribution using $10^5$ samples. The KL divergences are small, and quite similar to the corresponding KL divergences in Figure~7 of Ref.~\cite{BravoPrieto2022} obtained on the IBM {\tt ibmq{\_}santiago} device. The $s$ projection is better on \ibm\ with a KL divergence of 0.51, while the $t$ and $y$ distributions are slightly worse (KL divergences of 0.31 and 0.44, respectively). The second row of Figure~\ref{fig:ibmtorinoresults} displays the results for the two-dimensional correlations between the three distributions, while the bottom row displays the ratio between the reference samples ($10^5$ samples) and the samples generated by the style-based qGAN, where the white and light blue points signal regions of excellent agreement. The correlations are better captured on \ibm\ than on {\tt ibmq{\_}santiago}. Overall, the results on the IBM \ibm\ device show that the data augmentation is performed quite well and that circuit replication behaves as intended.

Our results for the data augmentation on the IonQ \ionq\ device are presented in Figure~\ref{fig:ionqaria1results}. We have explicitly deactivated debiasing, which is a standard mitigation option on \ionq, to keep the comparison focused on native unmitigated behavior. Similar to Figure~\ref{fig:ibmtorinoresults}, we show one-dimensional projections and KL divergences (top row), two-dimensional correlations (middle row), and ratios to reference samples (bottom row). The KL divergences are 0.25, 0.14, and 0.07 for $s$, $t$, and $y$, respectively. The correlation plots are also good and the ratio plots display a sizable amount of white and light blue points. Together with the IBM results on \ibm, the results show that the same style-based qGAN workflow can be executed with useful data-augmentation quality on both hardware modalities in this benchmark setup.

\begin{figure*}[t!]
  
  \hspace{1.2em}\subfigure[]{\includegraphics[width=0.30\textwidth]{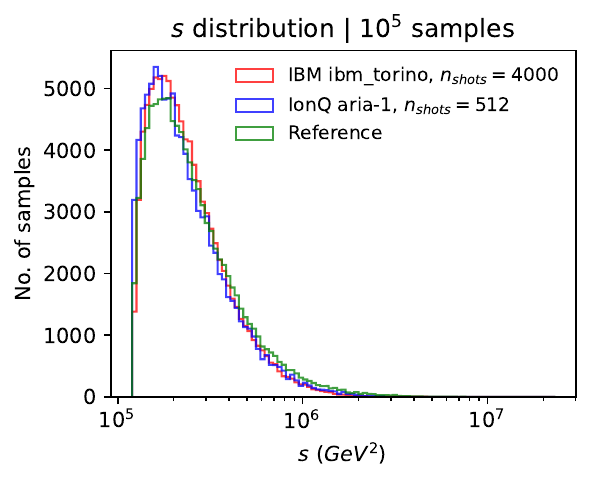}}%
  \subfigure[]{\includegraphics[width=0.30\textwidth]{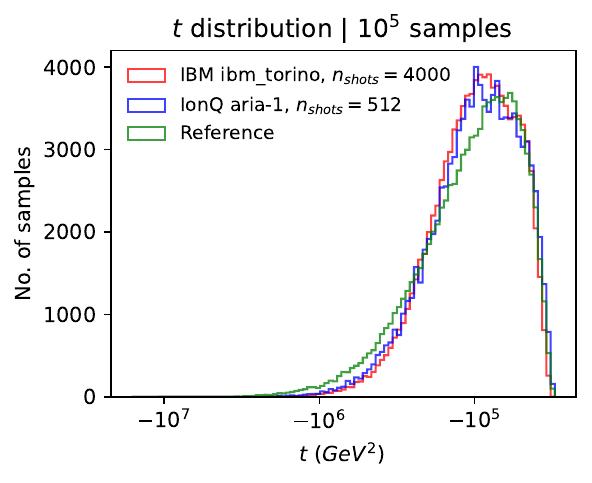}}%
  \subfigure[]{\includegraphics[width=0.30\textwidth]{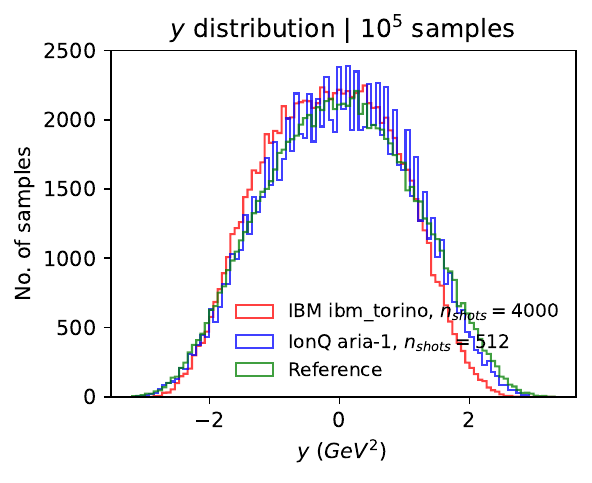}}

  \caption{\label{fig:ionqibmcompare}
    Comparison of the data augmentation on IBM \ibm\ (red lines) and on IonQ \ionq\ (blue lines) with the reference sample distribution (green lines).
    Panel (a): marginal $s$ distribution.
    Panel (b): marginal $t$ distribution.
    Panel (c): marginal $y$ distribution.}
\end{figure*}

We now compare the two hardware devices within the protocol limits stated above. The nominal KL divergences on \ionq\ are lower than on \ibm\ for all three marginals: 0.25 (0.14, 0.07) on \ionq\ versus 0.51 (0.31, 0.44) on \ibm\ for $s$ ($t$, $y$). The bottom rows of Figures~\ref{fig:ibmtorinoresults} and \ref{fig:ionqaria1results} also suggest somewhat better correlation capture on \ionq, especially for $y$, with more white and light blue points. Figure~\ref{fig:ionqibmcompare} provides a direct visual comparison of the one-dimensional marginal distributions. The red lines are associated with \ibm, the blue lines correspond to \ionq, and the green lines correspond to the reference data set. In the $y$ distribution, the blue line of \ionq\ is closer to the green reference line than the red line of \ibm. It should be noted that out of the 48 qubits on \ibm, the first 24 would be the best the machine can offer when transpiling the quantum circuits to the actual hardware while the remaining 24 are likely to be of lower quality. This contrasts with \ionq\ where only 24 qubits are used. However, we do not expect this difference to lead to a sizable difference in accuracy, as the comparison between the results on {\tt ibmq{\_}santiago} using three qubits~\cite{BravoPrieto2022} and the results on \ibm\ with 48 qubits shows no sizable difference in accuracy. Note that this benchmark is not a fully controlled physics-only comparison: shot counts differ, transpilation constraints differ, and qubit allocations differ between the two providers. The observed differences should, therefore, be interpreted as practical workflow-level outcomes for this task rather than universal hardware rankings.

It should be noted that only one run has been executed for each device. Ideally multiple runs would have to be executed in order to average the KL divergences over these multiple runs, but this would have required far more resources. However, we have estimated the error on the KL divergences by calculating the sample variance, which we denote here as $\vec{\sigma}$, on the expectation values of Equation~\ref{eq:samples} and using it as the exploration of the 68\% confidence-level interval on the KL divergences. We have generated 11 sets of $(s,t,y)$ distributions out of the post-processing of the sample vector $(x_i+ \delta_j \sqrt{\sigma_i})$, where $x_i$ and $\sigma_i$ are the components of the sample vector $\vec{x}$ generated by each run and the components of the corresponding sample variance vector $\vec{\sigma}$, respectively; $\delta_j = -1 + 0.2*j$, for $j$ running from 0 to 10. Selecting the maximal and minimal KL divergences amongst the 11 sets and comparing them to the nominal KL divergences, we have obtained, on the \ibm\ device, KL divergences of $0.51^{+0.10}_{-0.07}$, $0.31^{+0.07}_{-0.01}$, $0.44^{+0.01}_{-0.05}$ for $s$, $t$, and $y$ distributions, respectively. The same exercise on the \ionq\ device yields KL divergences of $0.25^{+0.10}_{-0.07}$, $0.14^{+0.09}_{-0.05}$, $0.07^{+0.06}_{-0.01}$ for $s$, $t$, and $y$ distributions, respectively. These errors do mitigate the difference in accuracy we observe between the IBM and the IonQ devices, such that the performance on the two devices are actually much closer than expected by looking at only the nominal values of the KL divergences. However, the IonQ device is still more accurate with lower KL divergences.

In order to get more insight into what could drive this difference in accuracy, we have also performed another data augmentation experiment on the IBM {\tt ibm{\_}cusco} device based on the Eagle chip, for which the two-qubit gate error rate is one order of magnitude larger than the error rate on the Heron chip in the \ibm\ device, as reported in Table~\ref{table:ibmcusco} in \ref{sec:appendixibm} compared to the specifications of \ibm\ presented in Table~\ref{table:specifications}. The results are presented in \ref{sec:appendixibm}, showing that overall the accuracy is quite similar, except for the correlation plots which are much better with \ibm\ than with {\tt ibm{\_}cusco}. These results, when compared to the results of the \ionq\ device, indicate that the two-qubit error rate has a moderate impact on the accuracy. However, the readout error rate is significantly different between IonQ devices and IBM devices, as presented in Table~\ref{table:specifications}. The one-order-of-magnitude improvement on IonQ \ionq\ compared to IBM \ibm\ is likely to explain the difference in accuracy that we observe.

It is quite well known that superconducting qubits have a readout error of the order of $1\%$ whereas the error rate for gate operations can be much lower, see e.g. Ref.~\cite{Li2023} where the single-qubit gate error rate falls below $10^{-4}$. There have been recent improvements in the readout error, see for example Ref.~\cite{Chen2023} where two-state readout fidelity reached $99.5\%$. The readout error is part of the more general SPAM error (state preparation and measurement error). A new interesting direction to significantly improve the SPAM error rate on trapped-ion qubits has emerged in the past few years, by replacing ytterbium ions with barium ions $^{137}\text{Ba}^+$. Experiments by IonQ have demonstrated that they can reach a SPAM error rate of only $0.04\%$\footnote{See \url{https://ionq.com/resources/state-preparation-and-measurement-with-barium-qubits}.} while Quantinuum has pushed the limit even further by reaching a SPAM error rate of $0.0096\%$~\cite{An2022}. We expect these radical improvements to have a significant impact not only on the accuracy of (style-based) qGANs, but also on the practical implementation of error correction which relies on repetitive mid-circuit measurements.

\subsection{Runtime, throughput, and practical guidance}
For this fixed workload, we summarize the practical implications of our measurements. Trapped-ion \ionq\ yields lower marginal KL in our setup, consistent with lower readout error and fewer SWAP overheads, at the cost of much longer wall-clock time per circuit. Superconducting \ibm\ yields shorter end-to-end campaign time and supports circuit batching, at the cost of somewhat higher marginal KL here. For similar shallow expectation-value generators and comparable protocols, practitioners who prioritize marginal distributional fidelity and can tolerate long campaign times may find trapped-ion execution aligned with their goals; those who prioritize turnaround and sample throughput may find superconducting execution aligned with theirs, optionally with readout mitigation on IBM stacks.

We have also recorded the timing of the runs. The timing on \ionq\ is much worse than on \ibm. This reflects the fact that trapped-ion qubits are significantly slower than superconducting transmon qubits as exemplified in Table~\ref{table:specifications}. The total number of jobs on \ionq\ is 12,500, equivalent to the total number of circuits executed on the device, for an average execution time per job of $17.3\pm 0.5$~s, as reported by IonQ. We have executed 512 shots on \ionq, so that the average execution time per circuit and per shot is $34\pm 1$~ms. The total execution time in order to obtain the full generated sample set on the IonQ quantum computer is 59~hrs, 57~min, and 31~s. 

On the IBM \ibm\ device, however, the total number of jobs is 22 as we can bundle multiple circuits in one job. We obtain an average QPU execution time per job and per circuit of $1.078\pm 0.006$~s, amounting to an average execution time per job and per circuit, taking into account the various pre- and post-processing steps of the results on the IBM cloud including the transpilation step, of $4\pm 3$~s. We have executed 4,000 shots on \ibm, so that the average QPU execution time per circuit and per shot is $0.269\pm 0.002$~ms and the average execution time per circuit and per shot is $1.1\pm 0.7$~ms. The total execution time in order to obtain the full generated sample set on the IBM quantum computer is 6~hrs, 43~min, and 47~s, out of which 1~hr, 52~min, and 11~s is on the QPU only.

Comparing the total time spent to obtain the full generated sample set, including the transpilation step, the IBM device is around 8.5 times faster than the IonQ device for the total execution. We should note, however, that the runs on \ibm\ have been executed with 48 qubits, a replication level of 16, requiring in total only 6,250 circuits, but deeper ones, while the runs on \ionq\ have been executed with 24 qubits, requiring in total twice the amount of circuits. Nonetheless, at the level of a single shot and single circuit execution, the difference is even larger for the execution time including transpilation: around 34~ms on the IonQ device against around 1~ms on the IBM device. This time difference is mitigated by the fact that we required many fewer shots on \ionq\ than on \ibm\ in order to obtain our results. Our results are a direct reflection of the sizable difference in the technical specifications reported in Table~\ref{table:specifications} regarding the time needed for gate operations and measurement readout.

\section{Conclusion}
\label{sec:conclusion}

Given the rapid evolution of NISQ hardware, controlled benchmark studies of fixed QML workloads are important for practical adoption. In this work we have performed a benchmark of a style-based qGAN data-augmentation workflow on two hardware modalities, superconducting transmon qubits and trapped-ion qubits.

Using the same dataset and a fixed trained generator, we execute the workflow on IBM \ibm\ and IonQ \ionq\ and report both quality and runtime metrics for the same set of real-world data as in Ref.~\cite{BravoPrieto2022} where the style-based qGAN algorithm was introduced, namely the $(s,t,y)$ distributions for $t\bar{t}$ production at the Large Hadron Collider. To the best of our knowledge, this is one of the first controlled style-qGAN hardware-to-hardware comparisons for this HEP data-augmentation task. Both quantum platforms deliver useful data-augmentation quality for the studied $(s,t,y)$ distributions using a shallow circuit, with low Kullback-Leibler (KL) divergences and reasonable correlation recovery. This was hinted at in Ref.~\cite{BravoPrieto2022} but a quantitative benchmark was lacking. In the native unmitigated setup used here, \ionq\ achieves lower KL values while \ibm\ achieves a significantly shorter end-to-end execution time. Our native comparison is performed as best as we can, owing to the fact that we do not have a perfect control of all operations behind the scene on each machine.

Compared to Ref.~\cite{BravoPrieto2022}, we have updated the execution stack and used circuit replication to improve sample throughput, using up to 24 qubits on \ionq\ and 48 qubits on \ibm\ for the same target sample count, $10^5$ samples from a training dataset of $10^4$ samples. This has led to a substantial reduction in end-to-end execution runtime.

The relative accuracy difference is consistent with lower readout error and all-to-all connectivity on the trapped-ion platform implying a smaller transpiled circuit and fewer errors accumulated until final measurement, while the speed advantage of IBM is consistent with faster gate and readout operations as well as circuit-batching capabilities. Indeed, while the two devices have similar two-qubit gate errors, the readout error rate on the IBM system is one order of magnitude higher than on the IonQ system. However, our conclusions remain bounded by benchmark limitations: one run per platform, different shot counts, and no per-backend retraining or user-level mitigation, the latter very likely to improve the results on IBM devices with simple readout error mitigation. It should also be noted that the results have not been averaged over multiple experiments for generating the $10^5$ samples due to quantum resources constraints. Such an average could statistically mitigate the observed difference in accuracy.

The execution time on \ibm\ is significantly shorter than on \ionq, driven by the significantly faster qubit operations on superconducting transmon qubits compared to trapped-ion qubits. In our runs, one circuit takes around 17~s on \ionq\ versus around 1~s on the \ibm\ QPU (around 4~s including transpilation and cloud overhead on IBM).

We expect that improving the readout error and the coherence time on the IBM Heron chip, as well as the newest optimized transpilation algorithm, should increase the accuracy of the results on superconducting transmon qubits, not to mention readout error mitigation. A shorter execution time is foreseen on the IonQ trapped-ion quantum computer when the gate operations become significantly faster and when the device architecture allows for running multiple circuits in parallel. It is worth noting that improving the readout error rates on IBM systems and the timing of gate operations on IonQ systems are also a requirement for reliable error correction which relies on repetitive mid-circuit measurement.

Future work should extend this benchmark in several directions. A benchmark campaign including a classical baseline generator under matched data-budget conditions could, for example, help comparing quantum and classical energy consumption. Repeated hardware campaigns would help reporting run-to-run statistics directly. While this work has performed a benchmark as close as possible to the native modality of the two platforms, a quantified mitigation- and error-suppression-aware versus unmitigated comparison on each platform, beyond simple readout error mitigation, could also shed light on the ultimate capabilities of each platform stack. Beyond these future directions for benchmarking, there are also several algorithmic improvements that could be fruitful such as general-purpose multi-programming methods, training the generator directly on the quantum hardware to absorb some of the hardware errors in the final trained parameters, or exploring hybrid training when only the last few epochs are trained on the quantum hardware.

\ack

The author wishes to thank Frederik Fl\"other and Rajiv Krishnakumar for fruitful discussions, as well as Cl\'ement Baglio for his support. The use of IBM Quantum services for this work, as well as the use of IonQ Compute services, is acknowledged. Discussions with the technical supports of IBM and IonQ are also acknowledged.

\section*{Supplementary Material}
No separate supplementary material is provided with this manuscript.

\section*{Author Declarations}
\paragraph{Conflict of interest\\}
The author has no conflicts to disclose.

\paragraph{Author Contributions\\}
Julien Baglio: Conceptualization, methodology, software, investigation, formal analysis, writing---original draft, writing---review and editing.

\section*{Data Availability}

The data supporting the findings of the study are available at the following URL:
\url{https://zenodo.org/records/11278538}. They allow to reproduce all the results displayed in the figures of the paper as well as the calculation of the errors on the KL divergences. The data supporting the calculation of the timing on the quantum computers cannot be made available in a format that is sufficiently accessible or reusable by other researchers.

\vspace*{0.5cm}

\appendix
\counterwithout{figure}{section}
\counterwithout{table}{section}
\renewcommand{\thefigure}{\arabic{figure}}
\renewcommand{\thetable}{\arabic{table}}
\setcounter{figure}{7}
\setcounter{table}{1}

\section{Noise simulation on IonQ}
\label{sec:appendixnoise}

\begin{figure*}[t!]
  \hspace{1.5em}\subfigure[]{\includegraphics[width=0.295\textwidth]{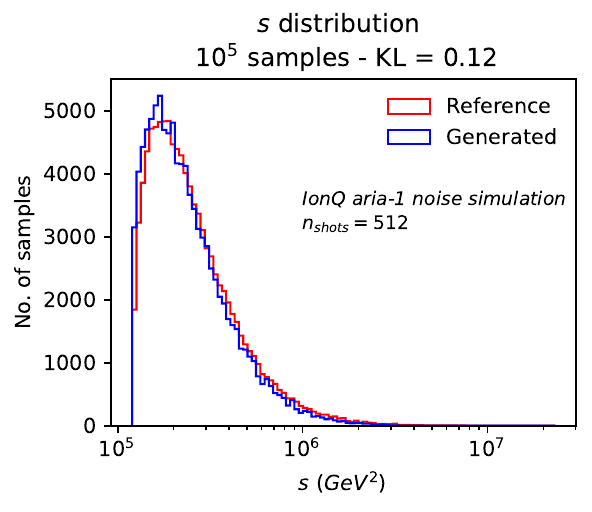}}%
  \subfigure[]{\includegraphics[width=0.295\textwidth]{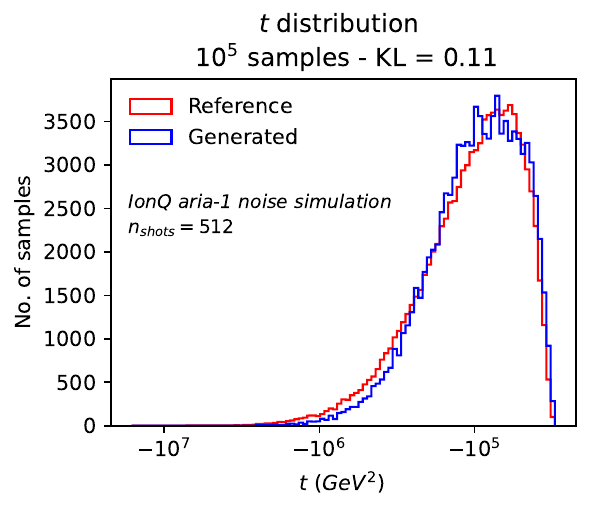}}%
  \subfigure[]{\includegraphics[width=0.295\textwidth]{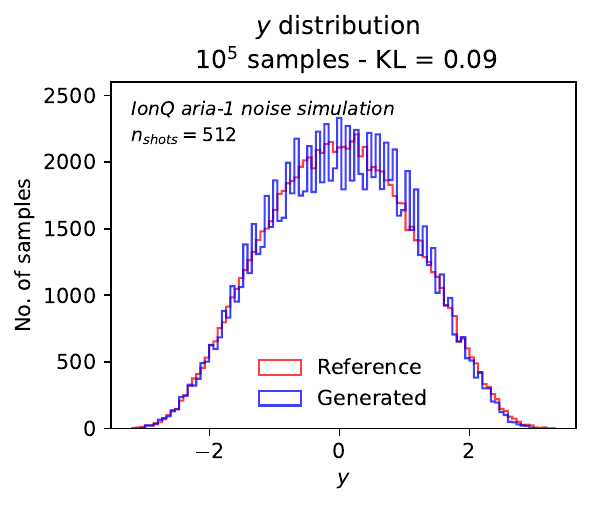}}

  \hspace{1.5em}\subfigure[]{\includegraphics[width=0.295\textwidth]{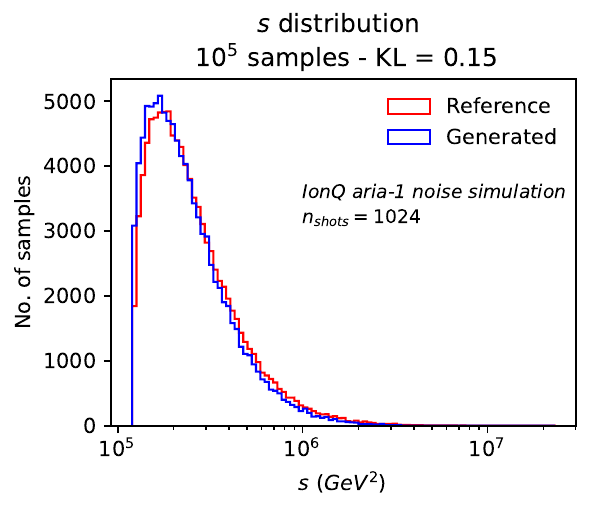}}%
  \subfigure[]{\includegraphics[width=0.295\textwidth]{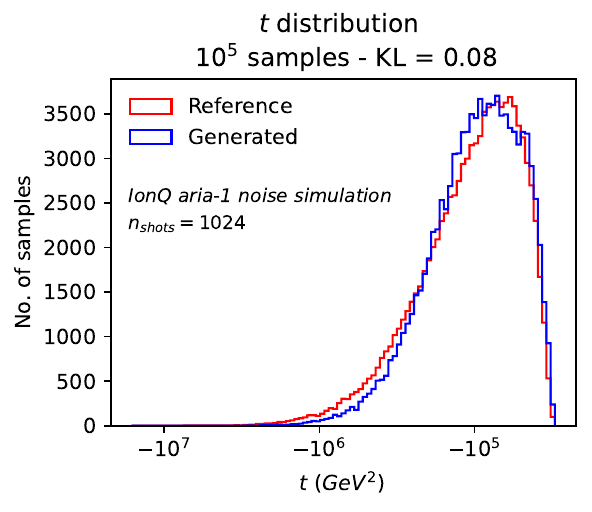}}%
  \subfigure[]{\includegraphics[width=0.295\textwidth]{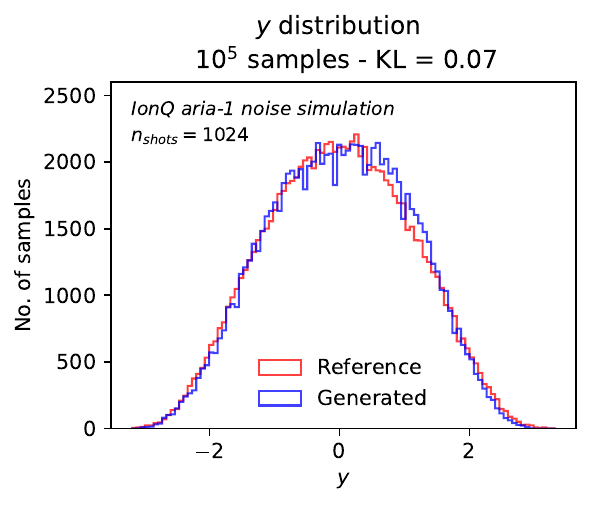}}
  
  \caption{\label{fig:arianoise1}
    Noise simulation of our data augmentation experiment using the style-based qGAN generator model trained with $10^4$ sample of Monte Carlo data for the physical observables $s,t,y$ in $pp\rightarrow t\bar{t}$ production at the LHC, using the noise model of the IonQ \ionq\ device.
    Panels (a)--(c): marginal distributions for 512 shots ($s$, $t$, $y$).
    Panels (d)--(f): corresponding marginal distributions for 1024 shots.
  }
\end{figure*}

\begin{figure*}[t!]
  \hspace{0.7em}\subfigure[]{\includegraphics[width=0.328\textwidth]{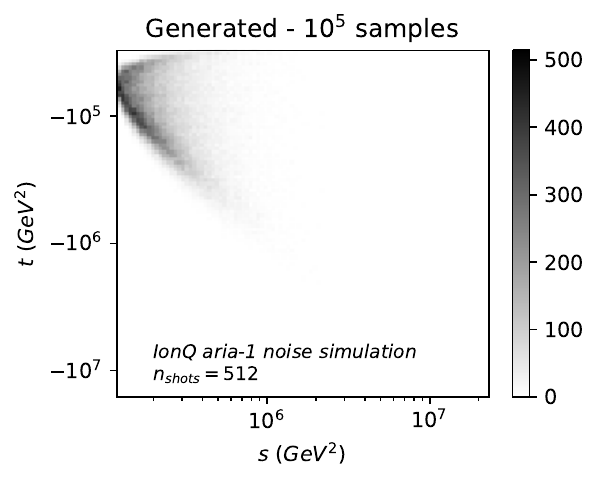}}%
  \subfigure[]{\includegraphics[width=0.303\textwidth]{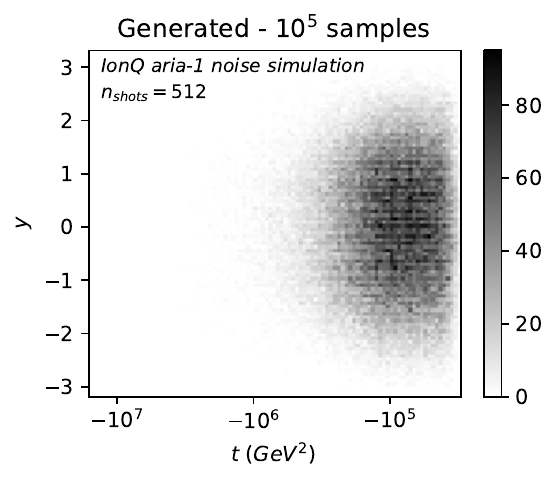}}%
  \subfigure[]{\includegraphics[width=0.318\textwidth]{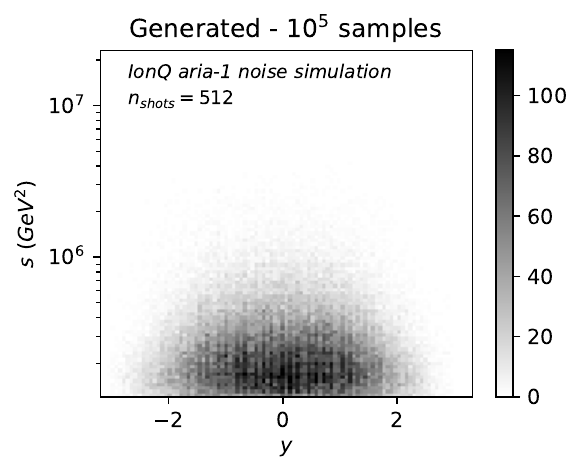}}

  \hspace{0.7em}\subfigure[]{\includegraphics[width=0.328\textwidth]{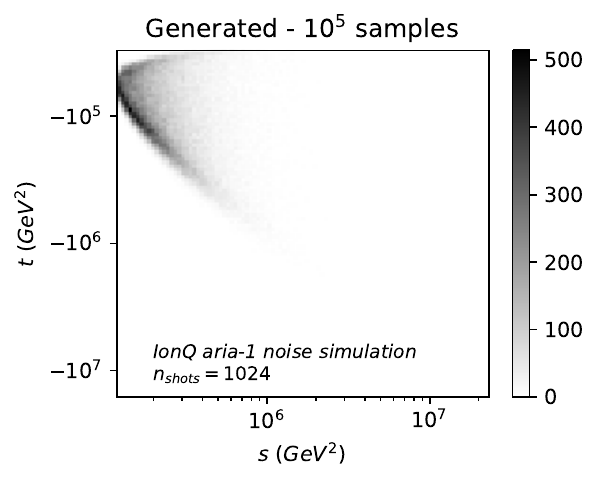}}%
  \subfigure[]{\includegraphics[width=0.303\textwidth]{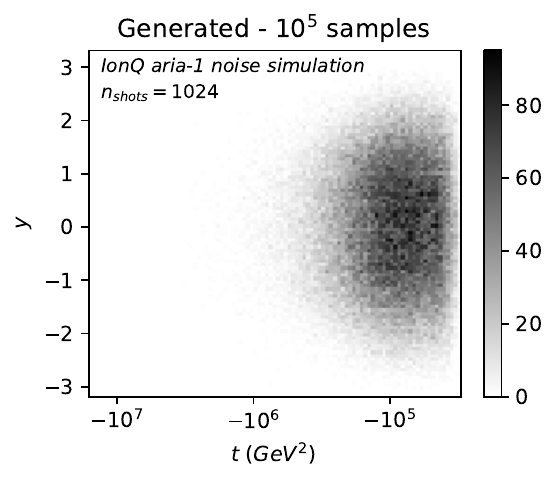}}%
  \subfigure[]{\includegraphics[width=0.318\textwidth]{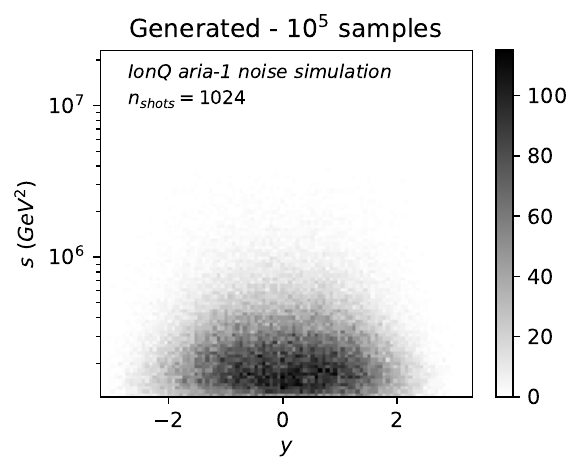}}

  \caption{\label{fig:arianoise2}
    Same as in Figure~\ref{fig:arianoise1} but for two-dimensional sampling projections.
    Panels (a)--(c): 512-shot correlations ($s$-$t$, $t$-$y$, $y$-$s$).
    Panels (d)--(f): corresponding 1024-shot correlations.}
\end{figure*}

\begin{figure*}[t!]
  \hspace{0.8em}\subfigure[]{\includegraphics[width=0.315\textwidth]{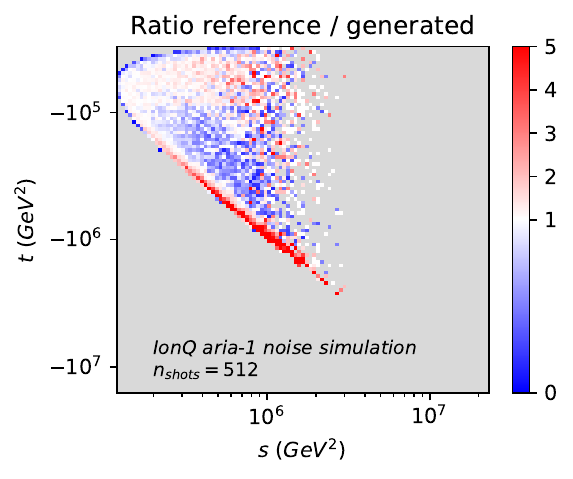}}%
  \subfigure[]{\includegraphics[width=0.299\textwidth]{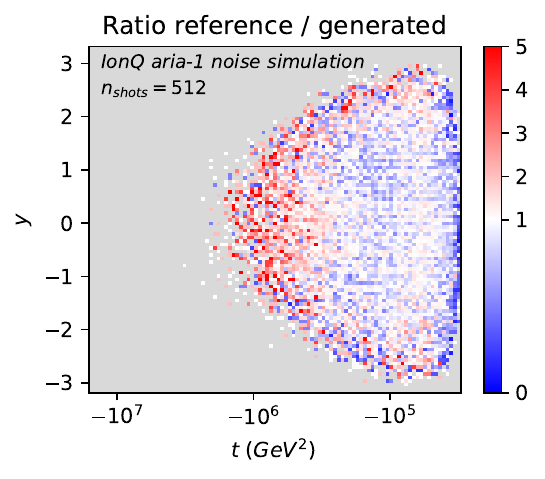}}%
  \subfigure[]{\includegraphics[width=0.310\textwidth]{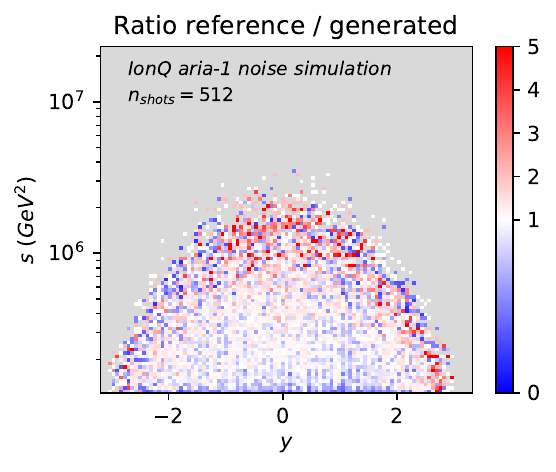}}

  \hspace{0.8em}\subfigure[]{\includegraphics[width=0.315\textwidth]{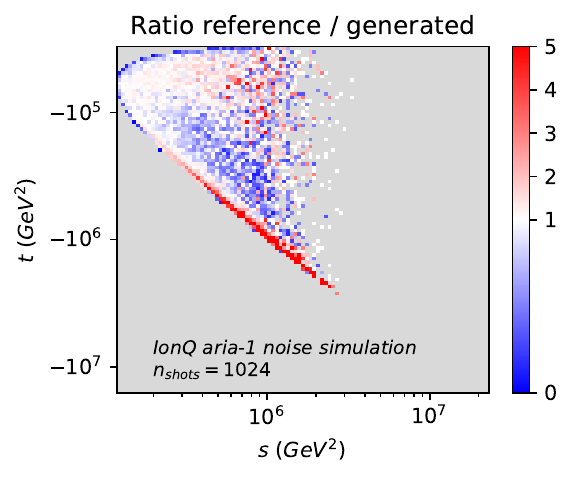}}%
  \subfigure[]{\includegraphics[width=0.299\textwidth]{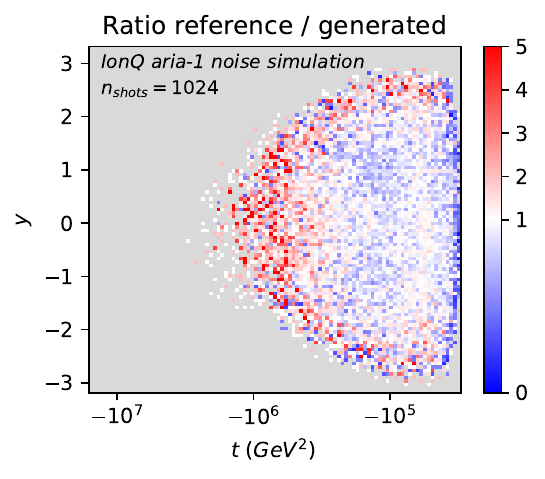}}%
  \subfigure[]{\includegraphics[width=0.310\textwidth]{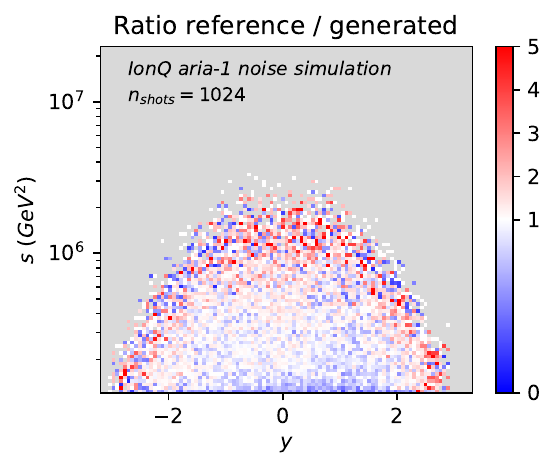}}
  
  \caption{\label{fig:arianoise3}
    Same as in Figure~\ref{fig:arianoise2} but for the ratio to the reference underlying prior Monte Carlo distribution.
    Panels (a)--(c): 512-shot ratio maps.
    Panels (d)--(f): corresponding 1024-shot ratio maps.
    A gray background highlights regions where reference and generated samples are identical.}
\end{figure*}

We have performed two simulations including the \ionq\ noise model in order to assess how many shots were required for the actual run on the IonQ hardware: with 512 shots and with 1024 shots. We present the results in Figures~\ref{fig:arianoise1}, ~\ref{fig:arianoise2}, and ~\ref{fig:arianoise3} for the marginal $(s,t,y)$ distribution, the corresponding two-dimensional sampling projections, and the ratio to the reference $10^5$ samples, respectively. In each figure the top row displays the results using 512 shots while the bottom row displays the results using 1024 shots.

The KL distances are not significantly different, the two-dimensional sampling distributions are very similar, and the ratios to the reference distribution display similar patches of white and light blue points, indicating that the behavior of the simulation does not significantly change from 512 shots to 1024 shots. The noise in the $y$ distribution for 512 shots is of a statistical nature and does not impact the results of the quantum generator, as indicated by the very similar KL divergences when comparing runs with 512 shots and 1024 shots.

Note that only one run has been performed for each choice of the number of shots, as was also done in the hardware runs. Statistics indicates that the typical uncertainty on the quantum measurement with 512 shots is around 8\%. To get confidence in the observation of the previous paragraph, we perform the same calculation of the sample variance as in Section~\ref{sec:results} for the hardware runs and we obtain, for $n_{shots}=512$, KL divergences of $0.12^{+0.05}_{-0.04}$, $0.11^{+0.03}_{-0.05}$, $0.09^{+0.01}_{-0.01}$ for $s$, $t$, and $y$ distributions, respectively. The same exercise for $n_{shots}=1024$ results in KL values of $0.15^{+0.04}_{-0.04}$, $0.08^{+0.04}_{-0.02}$, $0.07^{+0.06}_{-0.01}$ for $s$, $t$, and $y$ distributions, respectively. These errors do not change the picture of the previous paragraph and we have thus chosen to perform the runs on the actual quantum device using 512 shots to reduce the execution time.

\section{Results on the IBM Eagle {\tt ibm{\_}cusco} quantum system}
\label{sec:appendixibm}

We have also performed on October 7th, 2023, a run on the 127-qubit {\tt ibm{\_}cusco} device, based on the Eagle chip. The technical specifications given in Table~\ref{table:ibmcusco} and the comparison with Table~\ref{table:specifications} show that they are less good than on \ibm, in particular the one- and two-qubit gate error rates, while the readout error rate is comparable.

\begin{table}[t!]
  \centering
  \footnotesize
  \begin{tabular}{l|c}
    & {\tt ibm\_cusco} \tabularnewline
      \hline
      \# of qubits
    & $127$ \tabularnewline
      Coherence time $T_1$ ($\mu$s)
    & $131$ \tabularnewline
      Coherence time $T_2$ ($\mu$s)
    & $102$ \tabularnewline
      One-qubit gate time ($\mu$s)
    & $0.044$ \tabularnewline
      Two-qubit gate time ($\mu$s)
    & $0.487^a$ \tabularnewline
      Readout time ($\mu$s)
    & $4.00$ \tabularnewline
      Two-qubit gate error rate
    & $9.1\times 10^{-2}$ \tabularnewline
      Readout error rate
    & $5.7\times 10^{-2}$ \tabularnewline
    \hline
  \end{tabular}\\
  $^a$ECR gate.
  \normalsize
  \caption{\label{table:ibmcusco} List of the most important parameters from the technical specifications of the IBM {\tt ibm\_cusco} device using an Eagle chip (average values over all the qubits). The actual values change over time after each calibration of the system and reflect the specifications at the time of our experiment, on October 7th, 2023.}
\end{table}

\begin{figure*}[!htbp]
  
  \hspace{1.2em}\subfigure[]{\includegraphics[width=0.30\textwidth]{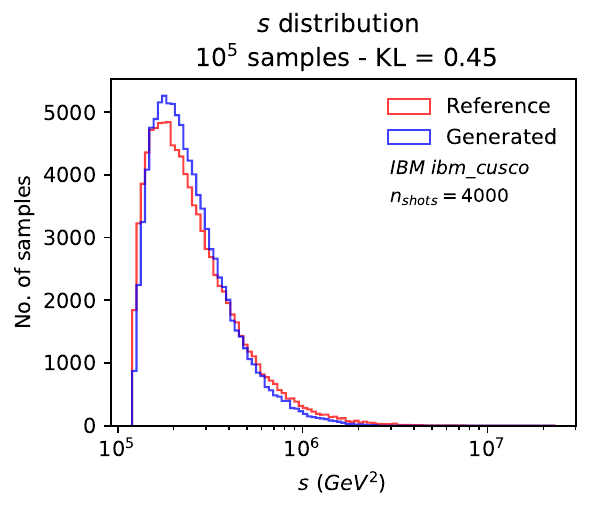}}%
  \subfigure[]{\includegraphics[width=0.30\textwidth]{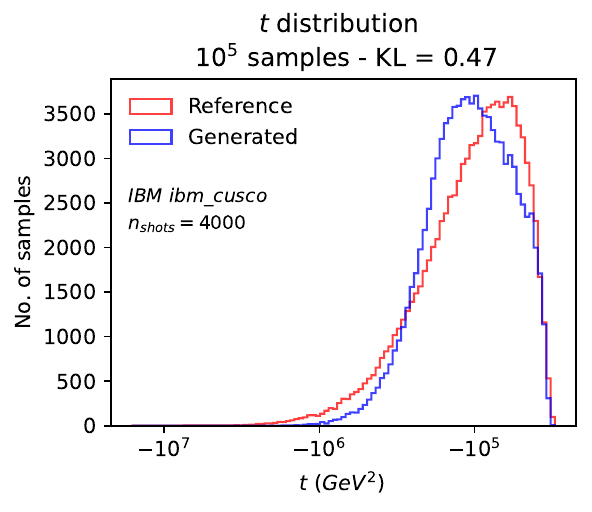}}%
  \subfigure[]{\includegraphics[width=0.30\textwidth]{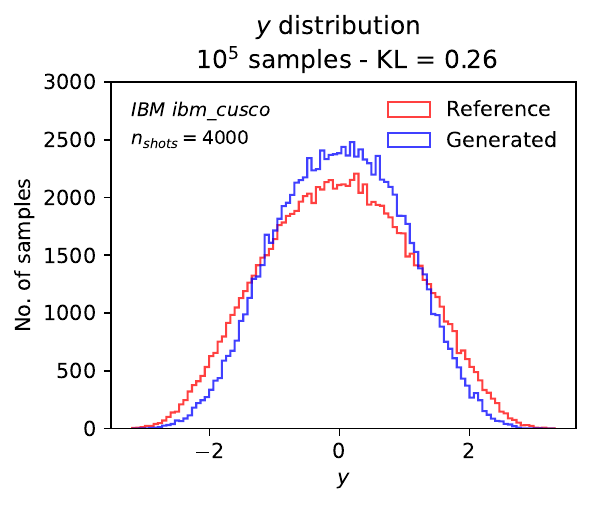}}

  \hspace{0.7em}\subfigure[]{\includegraphics[width=0.335\textwidth]{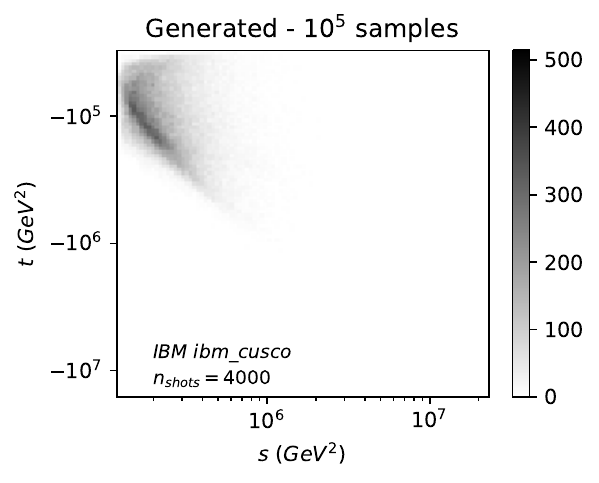}}%
  \subfigure[]{\includegraphics[width=0.310\textwidth]{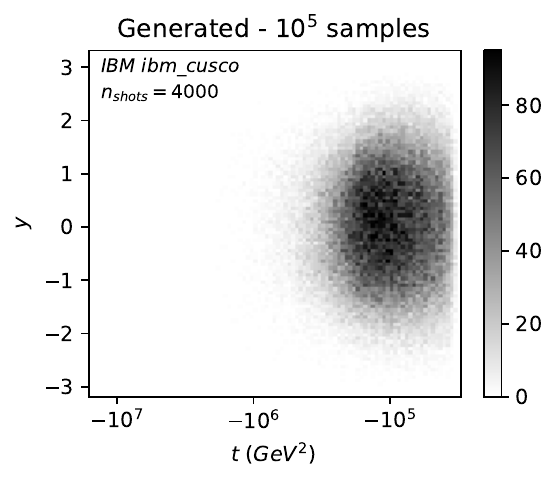}}%
  \subfigure[]{\includegraphics[width=0.325\textwidth]{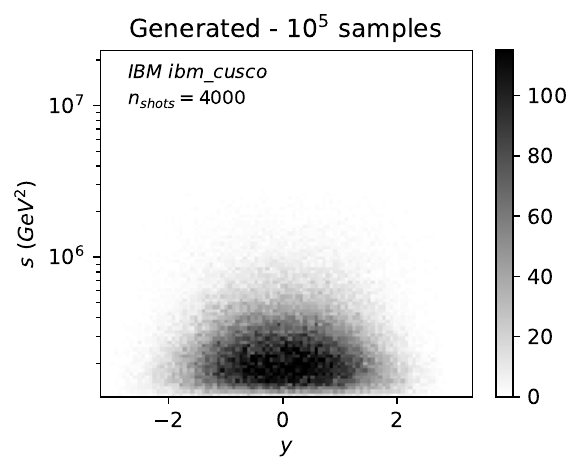}}

  \caption{\label{fig:ibmcuscoresults}
    Results for the data augmentation experiment on the IBM {\tt ibm\_cusco} device, using the style-based qGAN generator  trained with $10^4$ samples of Monte Carlo data for the physical observables $s,t,y$ in $pp\rightarrow t\bar{t}$ production at the LHC.
    Panels (a)--(c): marginal sample distributions for $s$, $t$, and $y$, respectively.
    Panels (d)--(f): corresponding two-dimensional sampling projections.}
\end{figure*}

The results are displayed in Figure~\ref{fig:ibmcuscoresults}. It is quite instructive to compare them to the results obtained with the \ibm\ device. The marginal distributions are not significantly different; it is better for the $y$ distribution on the {\tt ibm{\_}cusco} device, while the non-Gaussian $t$ distribution is better on the \ibm\ device. This observation indicates that the readout error rate, being the dominant source of error and comparable on both devices, drives the accuracy of the results on IBM systems as far as the marginal one-dimensional distributions are concerned. As the readout error rate is significantly lower on the IonQ \ionq\ device, the accuracy of the style-based quantum generator is better on the latter.

However, the comparison of the two-dimensional sampling projections does show a difference between {\tt ibm{\_}cusco} and \ibm. The \ibm\ device produces more accurate projections, as exemplified e.g. by comparing the $s-t$ projection produced by the {\tt ibm{\_}cusco} device (middle row of Figure~\ref{fig:ibmcuscoresults}) with the corresponding projection produced by the \ibm\ device (middle row of Figure~\ref{fig:ibmtorinoresults}). This reflects the one-order-of-magnitude improvement in the two-qubit gate error rate in the Heron \ibm\ device compared to the Eagle {\tt ibm{\_}cusco} device.

\clearpage
\bibliographystyle{iopart-num}
\bibliography{qgan_comparison_v2}

\end{document}